\documentclass[useAMS]{mn2e}
\input{psfig}
\usepackage{times}
\usepackage{bm}
\usepackage{amsmath}  

\def\Real{{\rm I\mathchoice{\kern-0.70mm}{\kern-0.70mm}{\kern-0.65mm}%
  {\kern-0.50mm}R}}

\def\bx#1{\leavevmode\thinspace\hbox{vrule\vtop{\vbox{\hrule\kern1pt
        \hbox{\vphantom{\tt/}\thinspace{\bf#1}\thinspace}}
      \kern1pt\hrule}\vrule}\thinspace}

\def\ps{{\bf s}}

\def\br{{\bf r}}
\def\bX{{\bf X}}

\def\be{\begin{equation}}
\def\ee{\end{equation}}
\def\ii{{\rm i}}
\def\ss{{\rm s}}
\def\oo{{\rm o}}
\def\ll{{\rm l}}
\def\mm{{\rm m}}
\def\ba{\begin{eqnarray}}
\def\ea{\end{eqnarray}}

\def\ltsima{$\; \buildrel < \over \sim \;$}
\def\lsim{\lower.5ex\hbox{\ltsima}}
\def\gtsima{$\; \buildrel > \over \sim \;$}
\def\gsim{\lower.5ex\hbox{\gtsima}}
\begin{document}
   \title[]{Sources of contamination to weak lensing three-point statistics: constraints from N-body simulations}
\author[]
{\parbox[]{6.in}{Elisabetta
    Semboloni$^{1}$\thanks{sembolon@astro.uni-bonn.de}, Catherine
    Heymans$^{2,3}$, Ludovic van Waerbeke$^{2}$, Peter Schneider$^{1}$ \\
  \footnotesize $^1$ Argelander-Institut f\"ur Astronomie, Auf dem H\"ugel 71,
  Bonn, D-53121, Germany. \\
$^2$ University of British Columbia, 6224 Agricultural Road, Vancouver, V6T 1Z1\, B.C.,Canada.\\
$^3$ Institut d'Astrophysique de Paris, 81bis Bd. Arago, F-75014, Paris, France.\\
}}
\maketitle
\begin{abstract}
We investigate the impact of the observed correlation between a galaxies shape and its
surrounding density field on the measurement of third order weak lensing shear statistics.
Using numerical simulations, we estimate the systematic error contribution to a measurement of
the third order moment  of the aperture mass statistic (GGG) from three-point intrinsic ellipticity
correlations (III), and the three-point coupling between the weak lensing shear experienced by
distant galaxies and the shape of foreground galaxies (GGI and GII).   We find
that third order
weak lensing statistics are typically more strongly contaminated by these physical systematics
compared to second order shear measurements, contaminating the measured three-point signal for
moderately deep surveys with a median redshift $z_\mm \sim 0.7$ by $\sim 15\%$.    It has been
shown that accurate photometric redshifts will be crucial to correct for this effect, once a
model and the redshift dependence of the effect can be accurately constrained.  To this end we
provide redshift-dependent fitting functions to our results and propose a new tool for the
observational study of intrinsic galaxy alignments.  For a shallow survey with $z_\mm \sim 0.4$ we
find III to be an order of magnitude larger than the expected cosmological GGG shear signal.
Compared to the two-point intrinsic ellipticity correlation which is similar in amplitude to the
two-point shear signal at these survey depths, third order statistics therefore offer a promising
new way to constrain models of intrinsic galaxy alignments. Early shallow data from the next
generation of very wide weak lensing surveys will be optimal for this type of study.
\end{abstract}
\section{Introduction}
Weak gravitational lensing  represents a powerful tool to investigate the
large-scale distribution of matter. The majority of lensing results to date
have focused on using two-point shear statistics to constrain  the matter density parameter
$\Omega_{\rm m}$ and the matter power spectrum normalisation $\sigma_8$
\cite{vWetal01,Hoetal02,Baetal03,Jaetal03a,Haetal03,Rhetal04,Heetal05,vWetal05,Maetal05,Seetal06,Hoetal06,Maetal07,Beetal07,Fuetal08}.
As these two parameters are strongly degenerate however there is great interest
in measuring higher order statistics as their combination with the two-point
 shear statistics can effectively break  the degeneracy   between
 $\Omega_{\rm m}$ and  $\sigma_8$ \cite{Beetal97}.
To date, there have been few measurements of  three-point shear statistics \cite{Beetal02,Jaetal03b,Penetal03}. These results found that the three-point shear statistics
were significantly affected by systematics
 even when the two-point statistics showed a very low systematic level.
The aim of this paper is to investigate, by using $\Lambda$CDM N-body simulations,
whether the intrinsic alignment of the sources (see for example Heavens et
al. 2000) and the correlation between the shear field and the intrinsic ellipticity of the sources
\cite{H04,Heetal06,Maetal06,Hietal07} can explain the presence of some systematics.

In the weak lensing regime the observed ellipticity $e_\oo$ of a source galaxy is
related to the original ellipticity $e_\ss$ through:
\be\label{gamma}
e_{\oo}\simeq e_\ss+\gamma,
\ee
where $\gamma=\gamma_1 + \ii \gamma_2$ is the complex shear, and $e$ is the complex
ellipticity defined as
\be\label{elli}
e =\frac{1-\beta^2}{1+\beta^2}\frac{{\rm exp}{(2 \ii \phi)}}{R},
\ee
where $\phi$ is the angle between the semi-major axis and the $x$-axis, $\beta$ is the ratio between the semi-major and semi-minor axis and $R$ is the response of a
galaxy to weak lensing shear field.
Note that in the commonly used  KSB method \cite{KSB}  the 
responsivity $R$ is expressed by the polarizability tensor which is computed for 
each object and relates the measured weighted ellipticity to the shear field.
In this analysis where no weight is applied we are able to  calculate the
average 
responsivity for our sample finding $R=0.89$ (see Rhodes et al. 2000,
Bernstein \& Jarvis 2002).

The two- and three-point ellipticity correlation functions between  the observed galaxies
$a$, $b$ and $c$ are given by :
\begin{eqnarray}
&\langle e_{\oo }^a e_{\oo }^b\rangle=\langle e_\ss ^a e_\ss ^b\rangle+ {\rm GI}  +\langle \gamma^a\gamma^b\rangle,\label{2pt}&\\
&\langle e_{\oo }^a e_{\oo }^b e_{\oo }^c\rangle=\langle e_\ss ^a e_\ss ^b e_\ss ^c\rangle+  {\rm GGI}  + {\rm GII}  +\langle \gamma^a \gamma^b \gamma^c\rangle,\label{3pt}&
\end{eqnarray}
where the GI, GII and GGI terms  are:
\begin{eqnarray}
&{\rm GI} =\langle e_\ss ^a\gamma^b\rangle+\langle e_\ss ^b\gamma^a\rangle,\label{GI}&\\
&{\rm GII}=\langle \gamma^a e_\ss ^b e_\ss ^c\rangle+\langle \gamma^b e_\ss ^c e_\ss ^a\rangle+\langle \gamma^c e_\ss ^ae_\ss ^b\rangle,\label{GII} &\\
&{\rm GGI}=\langle \gamma^a \gamma^b e_\ss ^c\rangle +\langle \gamma^b\gamma^c e_\ss ^a\rangle +\langle \gamma^c \gamma^a e_\ss ^b\rangle.\label{GGI} &
\end{eqnarray}
One assumes that galaxies are intrinsically randomly oriented so
the measurement of the observed ellipticity is an unbiased estimator of the
shear. In a similar way one assumes that  the measurement of the correlation between  
ellipticity of pairs and triplets of galaxies is an unbiased estimator of the
second and third order moments of the weak lensing shear field. 
 However, this assumption is not correct for galaxies that are physically close
as an intrinsic alignment  between the galaxies can be induced by a tidal field
which acts on cluster scales as has been observed by Brown et al. (2002) and Mandelbaum et al. (2006).
These results imply that the first term of Equation (\ref{2pt}), hereafter
referred as to  II, and the first term of Equation (\ref{3pt}), hereafter referred as to  III,  are non-zero and they systematically affect the value of the two- and three-point weak lensing 
shear statistics estimated through the observed ellipticity alignment.
It has been shown (King \& Schneider 2002, 2003, Heymans \& Heavens 2003, King 2005) that this bias can be removed
if close pairs of galaxies are downweighted in the lensing analysis, something which is possible only when
the redshift for each individual galaxy can be reliably estimated. For moderately
deep surveys such as CFHTLS Wide (Fu et al. 2008), the
intrinsic alignment does not  significantly affect  the estimation of the normalisation of power
spectrum $\sigma_8$. However, once the redshift
information is added it is possible to exploit all the information carried by the
shear signal, by using three-dimensional measurements. The measurement of
the two-point shear statistics in redshift bins, i.e. tomography (Semboloni
et al. 2006) and the 3D-lensing (Heavens et al. 2006, Kitching et al. 2007), are particularly
promising to constrain the equation of state of dark energy, but are also more
susceptible to stronger contamination by intrinsic alignment (Heymans et al. 2004, King 2005).  

A more subtle effect is the correlation between the induced weak lensing
shear and the foreground intrinsic ellipticity field, also called
``shear-shape'' correlation \cite{H04}.
This effect can be explained as follows.
Consider a background galaxy  with redshift $z^a$ whose light is deflected by
a foreground over-density at $z^b$.  Consider now a galaxy  at redshift $z^b$
stretched by the tidal forces generated by the dark matter over-density.  
Then the  term $\langle \gamma^a e_\ss ^b\rangle \neq 0 $. 
Similarly, for the three-point shear statistics, one can consider two sources with redshift $z^a$ and $z^b$  whose
light is deflected  by the same  halo hosting  a galaxy at redshift $z^c<z^a,z^b$: 
in this case the term $\langle \gamma^a\gamma^b e_\ss^c\rangle \ne 0 $.
Finally, one can imagine a background galaxy with redshift $z^a$ being sheared 
by an over-density hosting galaxies at redshifts $z^b$ and $z^c$ with the
result $\langle \gamma^a e_s^b e_s^c\rangle \ne 0 $. 

The first measurement of the shear-shape systematic effect on the Sloan Digital Sky Survey (SDSS) data 
by Mandelbaum et al. (2006) has been refined by Hirata et al. (2007), which uses
subsamples from the SDSS and 2SLAQ surveys to check the dependence of the
shear-shape correlation on the color, morphology and 
luminosity of galaxies. 
Hirata et al. (2007) predict that the  shear-intrinsic alignment coupling 
could  cause an underestimate of  the normalisation  of the power spectrum
$\sigma_8$ of up to ten percent.
Heymans et al. 2006 (hereafter H06) provide a simple toy model to investigate
this effect using  N-body simulations which we  describe in more detail in
section 3. This simple model also predicts  a shear-intrinsic alignment of
$10\%$ of the  expected weak lensing shear signal, and provides a good fit to the measurements of Hirata et al. (2007).
In contrast to the intrinsic alignment systematic that affects the lensing measured in the same tomographic redshift slice, the shear-shape systematic affects galaxies in very different tomographic redshift bins. The shear-shape systematic strongly reduces the precision  with which one is able to constrain the dark energy equation of state by tomography, as shown by Bridle \& King (2007).
 
Adopting the same strategy as H06, we analyse a set 
 of $\Lambda$CDM N-body simulations, in order to compare the amplitude of
 the intrinsic ellipticity and the shear-shape terms to the three-point weak lensing shear statistics.

The paper is organised as follows.
In the section 2 we define the quantities 
and we  describe  the  method used in this work to  measure  three-point statistics.
In  section 3  we describe the simulations  used in this work.
In section 4 we present measurements of the three-point intrinsic alignment, 
and in section 5 we show the evolution of  the three-point coupling between
shear and intrinsic ellipticity fields as a function of the redshift
distribution of the lenses and sources. We also provide a fitting formula for the three-point
shear-ellipticity correlation from different galaxy models. We conclude in
section 6.

\section{Three point  shear statistic measurement }
Following the approach of Pen et al. (2003) and Jarvis et al. (2004) we define
the complex `natural components' of the   three-point shear correlation
functions (Schneider \& Lombardi 2003) for a triangle of  vertex
$\bX_1$, $\bX_2$ and $\bX_3$ and separations vectors $\br_1=\bX_3-\bX_2$
$\br_2=\bX_1-\bX_3$, $\br_3=\bX_2-\bX_1$:
\begin{eqnarray}
\lefteqn
{\Gamma_0(r_1,r_2,r_3)=\langle \gamma(\br_1)\gamma(\br_2)\gamma(\br_3){\rm e}^{[-2\ii(\phi_1+\phi_2+\phi_3)]}\rangle,}\\
\lefteqn{
\Gamma_1(r_1,r_2,r_3)=\langle \gamma^\star(\br_1)\gamma(\br_2)\gamma(\br_3){\rm
e}^{[-2\ii(-\phi_1+\phi_2+\phi_3)]}\rangle,}\\
\lefteqn{
\Gamma_2(r_1,r_2,r_3)=\langle \gamma(\br_1)\gamma^\star(\br_2)\gamma(\br_3)
{\rm e}^{[-2\ii(\phi_1-\phi_2+\phi_3)]}\rangle,}\\
\lefteqn{
\Gamma_3(r_1,r_2,r_3)=\langle \gamma(\br_1)\gamma(\br_2) \gamma^\star(\br_3)
{\rm e}^{[-2\ii(\phi_1+\phi_2-\phi_3)]}\rangle,}
\end{eqnarray}
where $\gamma$ is the complex shear and we indicate with $\gamma^\star$ the complex conjugate of $\gamma$.
 The choice of the directions $\phi_i$ along  which one projects the shear are
  free as long as the value of $\Gamma_i$  does not depend on the triangles
 orientation. 
Assuming that the Universe is homogeneous  and isotropic, the eight real
 components of the shear correlation function depend on the side-lengths of
 the triangle, $r_1$, $r_2$, and $r_3$. Thus we prefer to use a derived statistic which is 
easier to visualize and interpret, namely the third order moment of the aperture mass statistic.

The complex aperture mass  is defined as :
\be\label{map}
M_\vartheta=M_{\rm ap}+\ii M_{\times}= \int d^2 \bm r Q_{\vartheta}(r) \gamma (\bm r)e^{(-2\ii \phi)},
\ee
where $Q_{\vartheta}(r)$ is a filter of characteristic size $\vartheta$.
Using the aperture mass statistics  allows one to uniquely separate the
E-mode ($M_{\rm ap}$) and B-mode  ($M_\times$) component   of the measured shear
(Crittenden et al. 2002), providing a powerful test for systematics. Indeed,
weak lensing shear fields are E-type fields; thus the presence of other sources 
of distortion can be revealed  by measuring   $M_{\times}$ statistics which are non-zero only for B-mode fields.

In practice, the variance and the third order moment  of the aperture mass  can be
measured  as a
function of  the two- and three-point shear statistics,  respectively \cite{Cretal02,Penetal02}.
 Estimating the third order moment of the aperture mass through three-point correlation
 functions is preferred however, as it yields a higher signal-to-noise ratio 
than  the direct measurement of the moments of 
the integral of Equation (\ref{map})
\cite{S02}. In addition, in the case of real data, where masked regions are
present, the estimate of the  aperture mass
statistic through the measurement of the correlation function
is  unbiased, whilst the direct measurement of the moments is biased.  
By choosing  the following filter,
\be\label{filter}
Q_{\vartheta}(r)=\frac{1}{2 \pi \vartheta^2}\Big(1- \frac{r^2}{2\vartheta^2} \Big) \exp \Big( -\frac{r^2}{2\vartheta^2} \Big ),
\ee
 the four components of  the third order moment \\ $\langle M_{\rm ap}^3\rangle$,
 $\langle M_{\rm ap}^2M_\times\rangle $,$\langle M_{\rm ap}M_\times^2\rangle$ and $\langle M_\times^3\rangle$   can be obtained
 through integration of  the three-point shear statistics  $\Gamma_i$ (see
 for example equations (45), (50) of Jarvis et al. 2004 and equations (61-71) of
 Schneider et al. 2005). There are other advantages of using aperture mass
 statistics defined by using Equation (\ref{filter}). This filter has
 infinite support but the exponential cutoff  allows one to assume
 a finite support for real calculations. Furthermore, the Fourier transformed
 filter,
\be\label{fftfilter}
I(\eta)=\frac{1}{2 \pi} \frac{\eta^4}{4} {\rm exp} (-\eta^2)
\ee
were we defined $\eta=s \vartheta$, is a very narrow window filter, probing essentially
modes with $s \sim  2/ \vartheta $.

We measure the four complex $\Gamma_i$ components using a binary tree code
built following the model suggested by Pen et al. (2003). We assign to each box
an  ellipticity $(e_1,e_2)$  and a position $(x,y)$ which are the average of
the ellipticity and positions of the galaxies contained in the box. To each
box we assign also  a weight which is the sum of the weights assigned to each
galaxy. 
The characteristic size $l_i$ of each box is chosen to be the distance between the
centre of the box and the furthest galaxy contained in the box. 
The correlation between triplets is computed  for triangle of sizes
$l_1$, $l_2$, $l_3$  satisfying the conditions:
\ba
\frac{l_1+l_2}{r_3} <0.1,~ ~\frac{l_2+l_3}{r_1} <0.1, ~ ~ \frac{l_3+l_1}{r_2} <0.1~.
\ea
Each triangle is ordered such that $r_3 \leq r_2 \leq r_1$. We measure $\Gamma_i$ for  triangles with $r_3$
between $5~\rm{arcsec}$ and $80~\rm{arcmin}$, ranging in  logarithmic bins of
width $\log(r_3)=0.1$. With this choice, we can assume that the distance between  each pair of galaxies
within the  boxes that we are correlating has a maximum
error of one bin.
Finally we integrate $\Gamma_i$ using equation (45) and equation (50) of Jarvis et al. (2004).
As we already anticipated, the filter  defined by  Equation (\ref{filter}) for a given size $\vartheta$ has
infinite support so we ideally
would need  to measure $\Gamma_i$ at all scales to perform the integral  giving the
components of the the third-order moment of the aperture mass.
However the integral is significant only for
triangles up to   $r_3\simeq 4 \vartheta $, implying that our measurement of the
third order moment is reliable up to an angular scale $\vartheta \simeq 20~ \rm{arcmin}$.
 We tested our algorithm against the direct measurement of the third-order moment
of the aperture mass on $\Lambda$CDM simulations in fields of $25~{\rm deg^2}$
and we found indeed a good agreement between the two methods at these scales.

Throughout the paper we  use only the aperture mass statistics and we call  
the systematic  produced by the intrinsic
alignments on the second and third order moments of the aperture mass II and III, respectively.
Similarly, we call GI the second order moment measured by using the filter
defined by the Equation (\ref{filter}) produced by the shear-shape coupling 
 and we call the third order moments GGI and GII.
Finally, we call  the second and third order moments of aperture mass produced by a weak
lensing field GG and GGG.

In order to know the importance of the GGI, GII and III when estimating
the third order moment of the aperture mass GGG using the observed ellipticity of galaxies
we have to compare those terms with  the third order moment  of the 
aperture mass produced by a weak lensing shear field. 
In the quasi-linear perturbation theory, the perturbation of the 
density field $\delta$ is considered
to be small so that it can be developed in a series
$\delta=\delta^{(1)}+\delta^{(2)}+....$, where $\delta^{(1)}$ is the linear
evolving density field and $\delta^{(n)} \propto {\mathcal O}((\delta^{(1)})^n)$. In
this approximation the $\langle M_{\rm ap}^3\rangle_{\vartheta}$ weak lensing 
signal is (Fry 1984, Schneider et al. 1998) :
\ba\label{map3}
 {\rm GGG}\equiv  \langle M_{\rm
   ap}^3\rangle_{\vartheta}=\frac{1}{2\pi}\frac{81H_0^6}{4c^6} \Omega_{\rm m}^3\int^{w_H}_{0} dw \frac{g^3(w)}{a^3(w)f_\kappa(w)}\\
\times\int_0^\infty d^2 \ps_1 P\Big(\frac{s_1}{f_\kappa(w)},w\Big)I(s_1\vartheta)\nonumber\\
\times \int_0^\infty d^2 \ps_2 P\Big(\frac{s_2}{f_\kappa(w)},w\Big)I(s_2\vartheta)\nonumber\\
\times I(|\ps_1+\ps_2|\vartheta) F_2(\ps_1,\ps_2),\nonumber
\ea
with 
\be\label{distr}
g(w)=\int_w^{w_H} dw^\prime p_s(w^\prime) \frac{f_\kappa(w-w^\prime)}{f_\kappa(w^\prime)}.
\ee
$f_\kappa(w)$ is the comoving angular diameter distance, $P(s,w)$ is the 3
dimensional  power spectrum of matter fluctuations, $p_s(w)$ is the comoving
distance distribution  of the sources, $I(s\vartheta)$ is the Fourier transform
of the
filter as defined by Equation (\ref{fftfilter}) and $F_2(\ps_1,\ps_2)$ is  the
coupling between two different modes  of density fluctuations characterised by
the  wave vectors $\ps_1$ and $\ps_2$. We compute $F_2(\ps_1,\ps_2)$ using  the fitting formula suggested by Scoccimarro \& Couchman (2001).

\section{N-Body simulations and galaxy models}
\begin{figure*}
\begin{tabular}{cc}
\psfig{figure=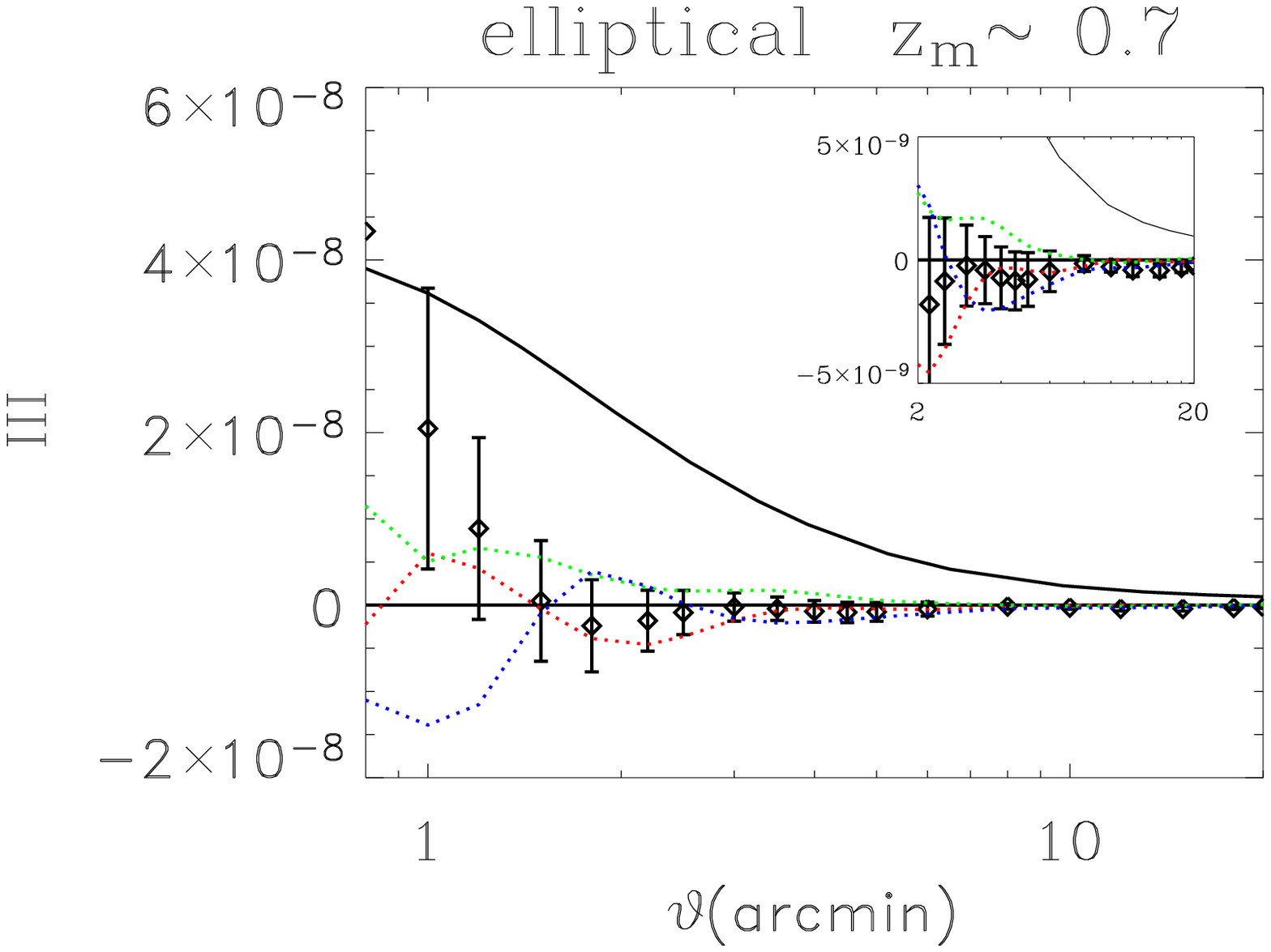,width=.45\textwidth} &\psfig{figure=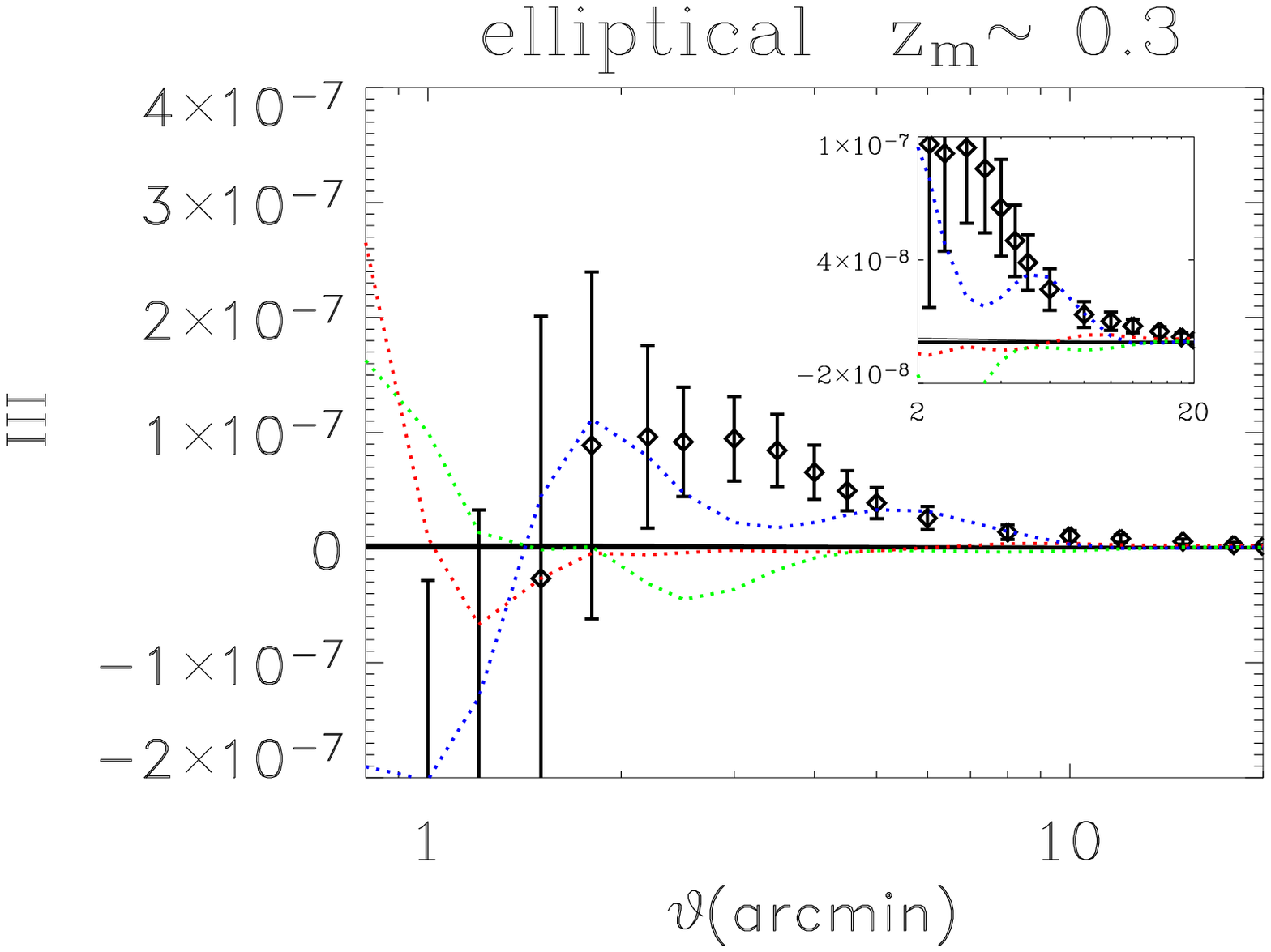,width=.45\textwidth}\\
\psfig{figure=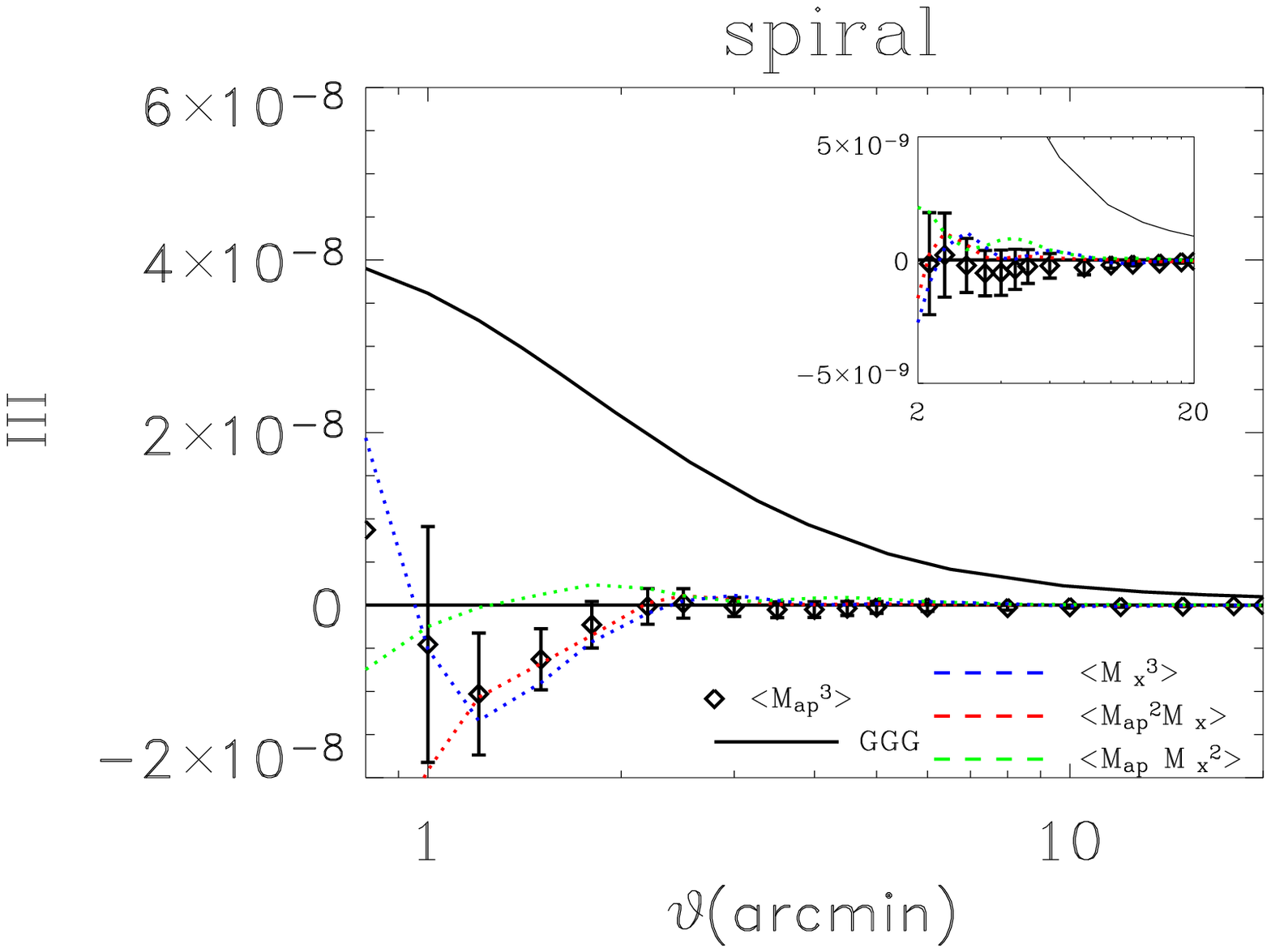,width=.45\textwidth} & \psfig{figure=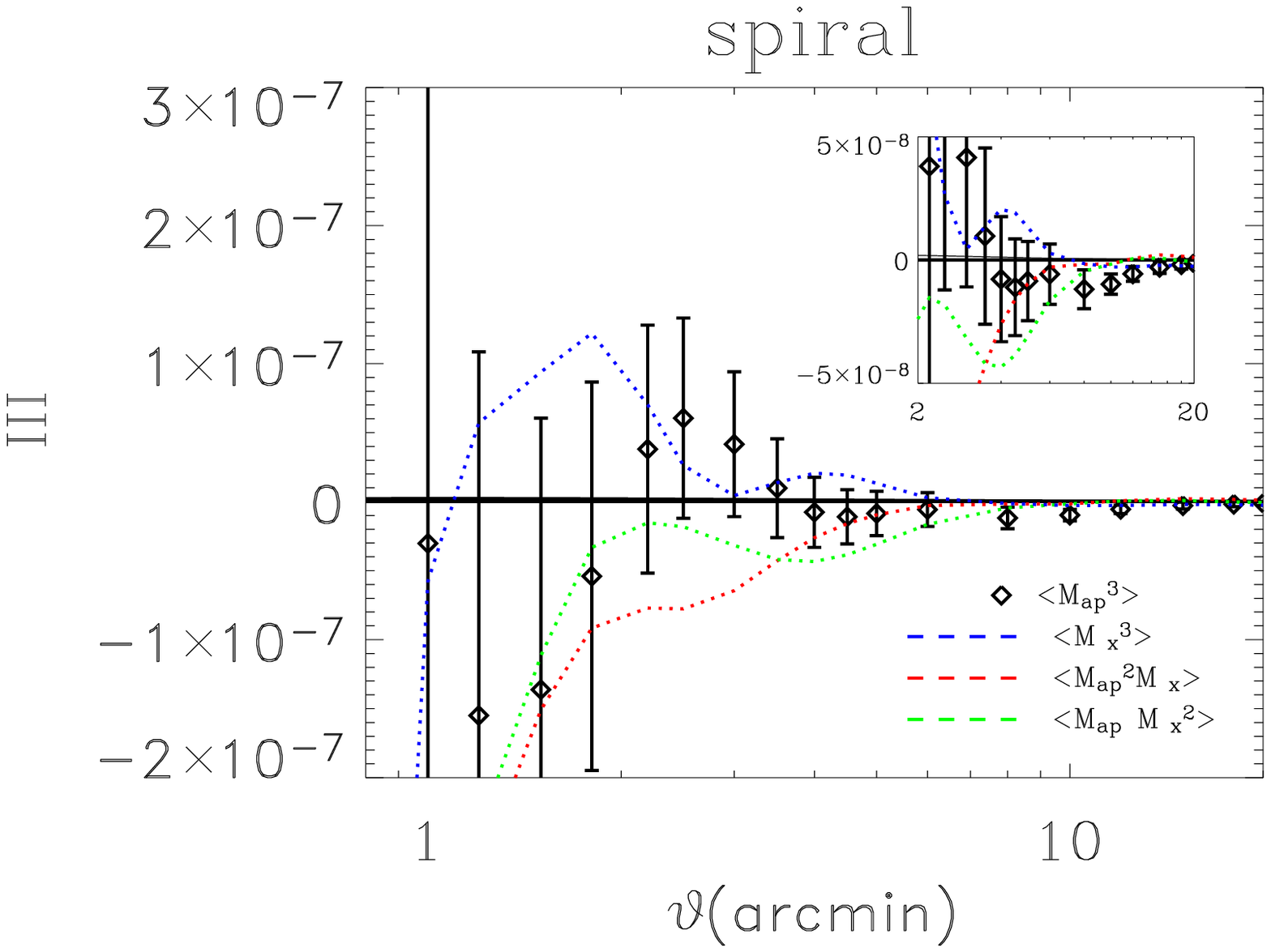,width=.45\textwidth}\\
\psfig{figure=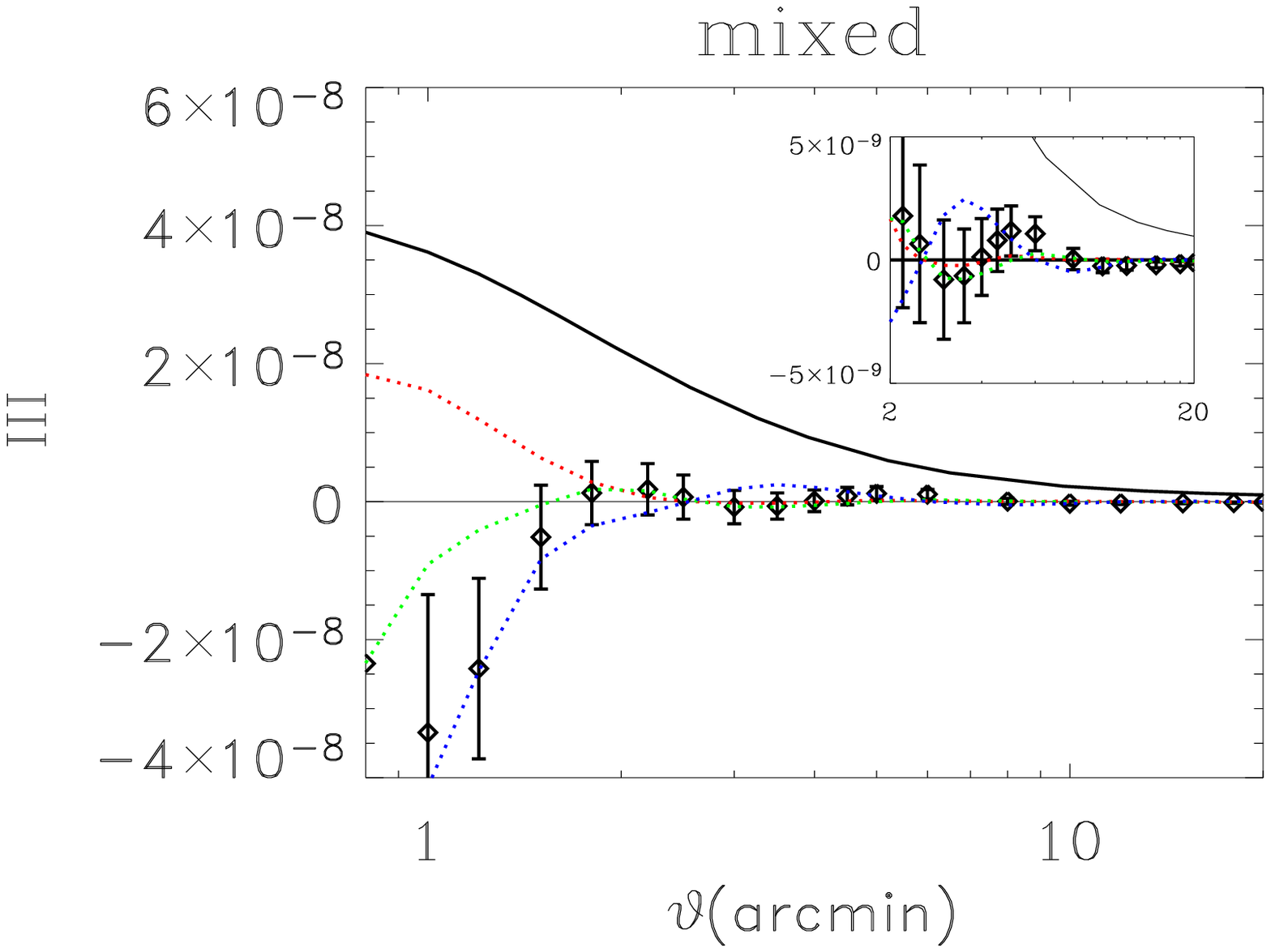,width=.45\textwidth} & \psfig{figure=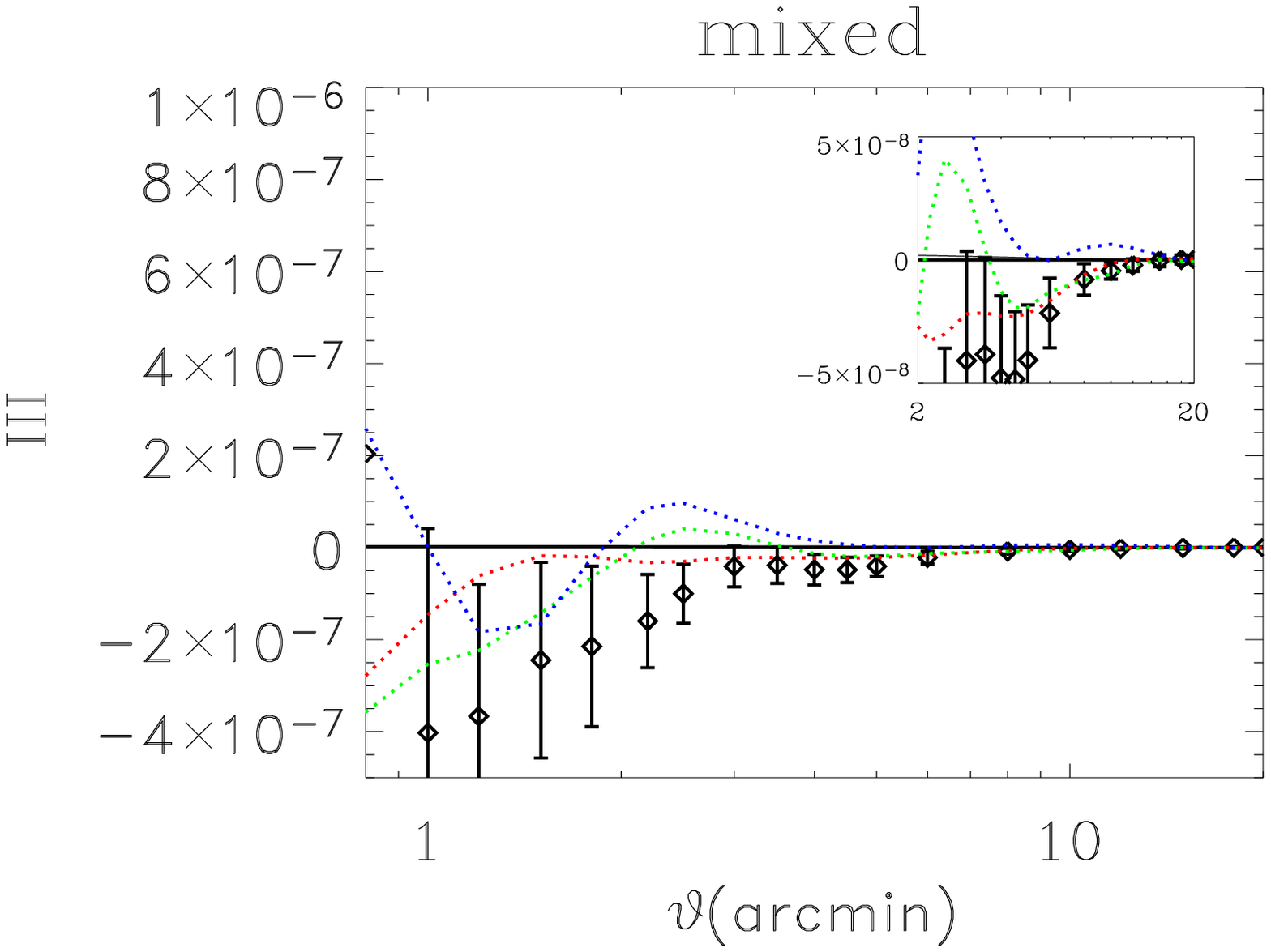,width=.45\textwidth}
\end{tabular}
\caption{\label{eee_res} Left panel:  $\langle M_{\rm ap} ^3\rangle$ signal 
  generated by the  intrinsic alignments (III) as a
  function of the angular size $\vartheta$.
  By comparing the E-modes (black diamond) with the expected shear signal (black solid line) one can determine
  the level of contamination produced by the existence of intrinsic alignments
  between galaxies. The shear signal has been computed using Equation
  (\ref{map3}) for the same  redshift distribution used to compute the
  intrinsic alignment term III. The median redshift of this 
  distribution is $z_\mm \sim 0.7$.
 The error-bars are  the dispersion of the average signal measured in twelve
  independent realizations. The B-mode signal, $\langle M_\times^3\rangle$, 
 and the E/B components,  $\langle M_{\rm ap}^2M_\times\rangle $ and $\langle
  M_{\rm ap}M_\times^2\rangle$, are shown
  dashed and for clarity we chose not to plot the error bars. The right panel
  shows the same results  as the left panel  for
  a lower redshift sample of galaxies  $0<z<0.4$. 
For this redshift distribution
 characterised by  $z_\mm \simeq 0.3$, the predicted shear signal GGG $\sim
  10^{-9}$ is  indistinguishable from  zero in this plot. }
\end{figure*}
In this analysis we use the same set of simulations used as in H06. We recall here only the main characteristics and refer the
reader to H06 for more detailed information.
The N-body simulation is a box of $300 h^{-1} {\rm Mpc}$ realized  using a
$\Lambda$CDM cosmology. The matter density parameter is
$\Omega_{\rm m}=0.3$, the cosmological constant is $\Omega_\Lambda=0.7$, 
the normalisation of the power spectrum of matter fluctuations is
$\sigma_8=0.8$, the reduced Hubble constant is $h=0.7$ and the baryon
density parameter is  $\Omega_{\rm b} h^2=0.2$. 
The simulation started at redshift $z=60$ and  evolved to $z=0$.
Twelve lines of sight were built by stacking boxes back along the $z$-axis
with random shifts  between the boxes in the $x$ and $y$ direction in order
to avoid artificial correlations, producing (almost) independent realisations.

The over-densities are identified using a `friends of friends' group finder,
which allows one to identify the halos. The particle mass of $1.7 \times
10^{10} ~h^{-1} \rm {M_\odot}$  allows us to find bound halos with masses
larger than  few $10^{11} ~ h^{-1} \rm {M_\odot}$. 
The halos are then populated with
galaxies. The luminosity is assigned following the  conditional luminosity
function (CLF) of Cooray \& Milosavljevic (2005) and the ellipticity is assigned
using a toy model that has been shown  effective at reproducing  the
observations of Mandelbaum et al. (2006) and Hirata et al. (2007). 
In this model, elliptical galaxies  are given the
same ellipticity as their parent halos. Spiral galaxies are modeled as a thick disk
oriented almost perpendicular to the angular momentum vector of the halo  with
a mean random misalignment of $\sim 20 ~\rm{deg} $  (van den Bosch et al. 2002, Heymans et al. 2004). At each identified dark matter halo in the simulations we generate both a spiral and an elliptical galaxy so that we
can investigate the results as a function of  morphology. 
We define the ellipticity of the model galaxies using Equation (\ref{elli}).  The resulting ellipticity distribution has a zero mean and dispersion of $\sigma_e=0.33$.  

In the results that follow we present three models: one composed exclusively of  spiral galaxies, one composed exclusively of  elliptical galaxies and one containing both
the  morphologies denoted `mixed'. The mixed model has been built
following Cooray \&  Milosavljevic (2005) and it contains  $\sim 30\%$
elliptical galaxies.
The mixed model is the most realistic model out of the three tested; the GI and II components predicted using the mixed model agree with  the ones
measured using the SDSS data (Mandelbaum et al. 2006, Heymans et al. 2006, Hirata et al. 2007). 
The number density of the final catalogue, which contains galaxies between
$0<z<1$ and is complete up to $r = 25.5 $, is  about 5 per $ \rm {arcmin^2}$, which is
considerably lower than what is expected for such a survey. This is due on one
hand to the fact that the low resolution of the simulations implies the loss of
low-mass halos (only halos with mass larger than  few $10^{11} ~ \rm
{M_\odot}$ are identified) and that each halo is  populated  with only
one galaxy. The lack of the  low-mass halos and satellite galaxies is the
main limitation of our model and this point is discussed further in the
conclusions.

Finally, these same N-body simulation which have been 
populated with galaxies are ray-traced to 
produce twelve projected mass distributions $\kappa$ for a source plane at
$z_\ss= 0.45$ and
for $z_\ss=1.05$.  We use these simulated $\kappa$ maps 
in section 5 to study the shear-shape effect. 
Each projected mass  map covers an area of $5\times 5 ~\rm{deg^2}$ in 
$2048\times 2048$ pixels. 

\section{Intrinsic ellipticity alignment}

In this section we report the third order moment  
components of the aperture mass given
by the intrinsic alignment of the sources.
We measure the  third order moment 
of the aperture mass for a survey with maximum depth $z \sim 1 $ and 
compare the measured intrinsic alignment to the
 expected weak lensing   GGG signal  from Equation (\ref{map3}). 
The redshift distribution
 of the sources $p_\ss(w)$ is computed directly  from the catalogues used
 for the III measurement so that both III and GGG terms have been computed for
 the same survey.

Figure \ref{eee_res}  shows  the three-point intrinsic alignment for our three
toy models: elliptical (upper panel), spiral (middle panel) and mixed (lower
panel) for two survey types; one with a median redshift of $z_\mm \sim 0.7$ (left panels) and the other with $z_\mm \sim 0.3$ (right panels). 
The error-bars on the measurements have been computed as the dispersion in the average signal measured in the twelve realisations, 
thus they include statistical and cosmic variance.  
For the moderately deep survey ($z_\mm\sim 0.7$) the III term is consistent with zero for all the three models.
 This result is in agreement with H06 who find that the two-point  intrinsic ellipticity
 term II, is consistent with zero for the same survey characteristics.
As expected, however, we find that  when the redshift depth of the survey  decreases the III term
becomes significant due to the fact that   for  lower redshift galaxies, a given angular distance  corresponds to a physically closer triplet.
Comparing now the results for the three different toy models for this shallow survey we find very different results.
 Figure \ref{eee_res} shows that the III term is  significantly positive for  $\vartheta > 2$ arcmin for elliptical galaxies  (upper right panel)  meanwhile it is slightly negative  for the spiral model  (middle right panel) and significantly  negative  for the mixed model  (lower right panel).  
The mixed model
contains roughly $70\%$ of spiral galaxies and $30\%$ of elliptical galaxies,
such that the
most frequent triangle  in the three-point measurement contains two spiral galaxies and one elliptical galaxy. As this type of
correlation does  not exist when one considers only elliptical or only spiral
morphologies, the mixed result is not a weighted average of the result
containing only elliptical or spiral galaxies. We verified that  the third order moment of  the aperture
mass is indeed negative when we correlate triplets containing two spirals and
one elliptical.
Figure \ref{eee_res} shows that for shallow surveys ($z_\mm \lsim 0.5$) the intrinsic 
alignment dominates the signal and suggests that  also for deeper surveys   the intrinsic
alignment could affect the measured  the three-point weak lensing signal. 

\begin{figure}
\psfig{figure=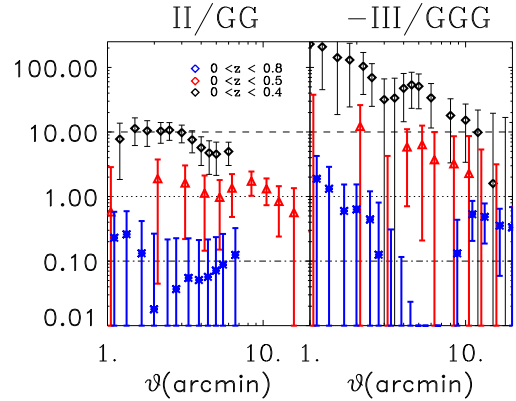,width=0.45\textwidth}
\caption{\label{GGvsII} Left side: ratio between the
intrinsic alignment (II) and the aperture mass variance 
(GG) for three  surveys. Right side:
ratio between III
  the expected weak lensing signal (GGG)  for the same surveys as the  left
  panel. The galaxy model is a  mixed model. The expected second and third
  order aperture mass moment are  computed for each survey by using the same
  source redshift distribution used to measure the III and II terms.
  These distributions are characterized by: $z_\mm\sim 0.31$ for source distribution
  between $0<z<0.4$,  $z_\mm \sim 0.4$  for source distribution $0<z<0.5$ and
  source distribution $z_\mm \sim 0.6$ for $0<z<0.8$. }
\end{figure}

In order to quantify the effect of the intrinsic alignment on the 
 weak lensing shear statistics we compare the amplitude
of the two- and three-point intrinsic alignment signal II and III with the
 expected  weak lensing signal, respectively GG and GGG,  
 for  different shallow surveys. Figure \ref{GGvsII} shows the amplitude
of the II/GG (left panel) and III/GGG (right panel) ratios  for three
 different surveys depth using  the mixed
model as an example.
We find that for a given survey depth, the ratio between the III term and expected weak lensing third order moment  GGG,
 is generally
higher than the ratio between the II term and the variance of the weak lensing
aperture mass GG.
One can think to reduce significatively the intrinsic ellipticity
contamination to the two- and three-point shear statistics by removing
very low redshift sources ($z \lsim 0.2$ ) which contribute to the
intrinsic alignment signal but not to the weak lensing signal.  We checked that
the
ratio II/GG and III/GGG drops on small scales $\vartheta \lsim 10$ arcmin  if
galaxies with $z<0.2$ are rejected, but it is almost unchanged at larger scales.

Figure \ref{eee_res} and Figure \ref{GGvsII} show  that without  a technique to correct for the intrinsic
 alignment it will be  impossible have a precise measurement of the three-point shear statistics in 
shallow surveys ($z_\mm \lsim 0.5$). For moderately deep surveys like the CFHTLS Wide, the
ratio III/GGG is consistent with zero as shown by the left panels of Figure
\ref{eee_res}. However, for larger surveys, where the statistical and cosmic
 variance are small   the intrinsic alignment could still play a role in the
 precision with which one can constrain the cosmological parameters.  
This can be seen in Figure \ref{GGvsII} where 
for a relatively deep survey ($z<0.8$) the
 intrinsic alignment contribution is likely to be  non-zero, specially for
 the third order moment of the aperture mass.

This result strongly  supports the need for 
reliable photometric redshift for weak lensing studies.
 Moreover,  Figure \ref{GGvsII} demonstrates  that if the redshift is known, one can select sources in order
to enhance the II and III signal relative to the contribution  
of the GG and GGG terms to the measured  two and three-point shear statistics.
This is an important result since one needs a good intrinsic 
alignment  model to be able
to remove this contribution   from the two-point (three-point) weak lensing
shear statistics (King 2005). The fact that the III/GGG signal is  more 
significant than the II/GG
signal implies that it may be easier to study intrinsic alignments 
using three-point statistics. 
In this respect Figure \ref{GGvsII} shows that large
multi-band surveys like SDSS and KIDS or PanSTARRS, represent
 excellent surveys to investigate  and model the intrinsic alignment of
elliptical and spiral galaxies.

 \section{Shear ellipticity alignment }

\begin{figure*}
\begin{tabular}{cc}
\psfig{figure=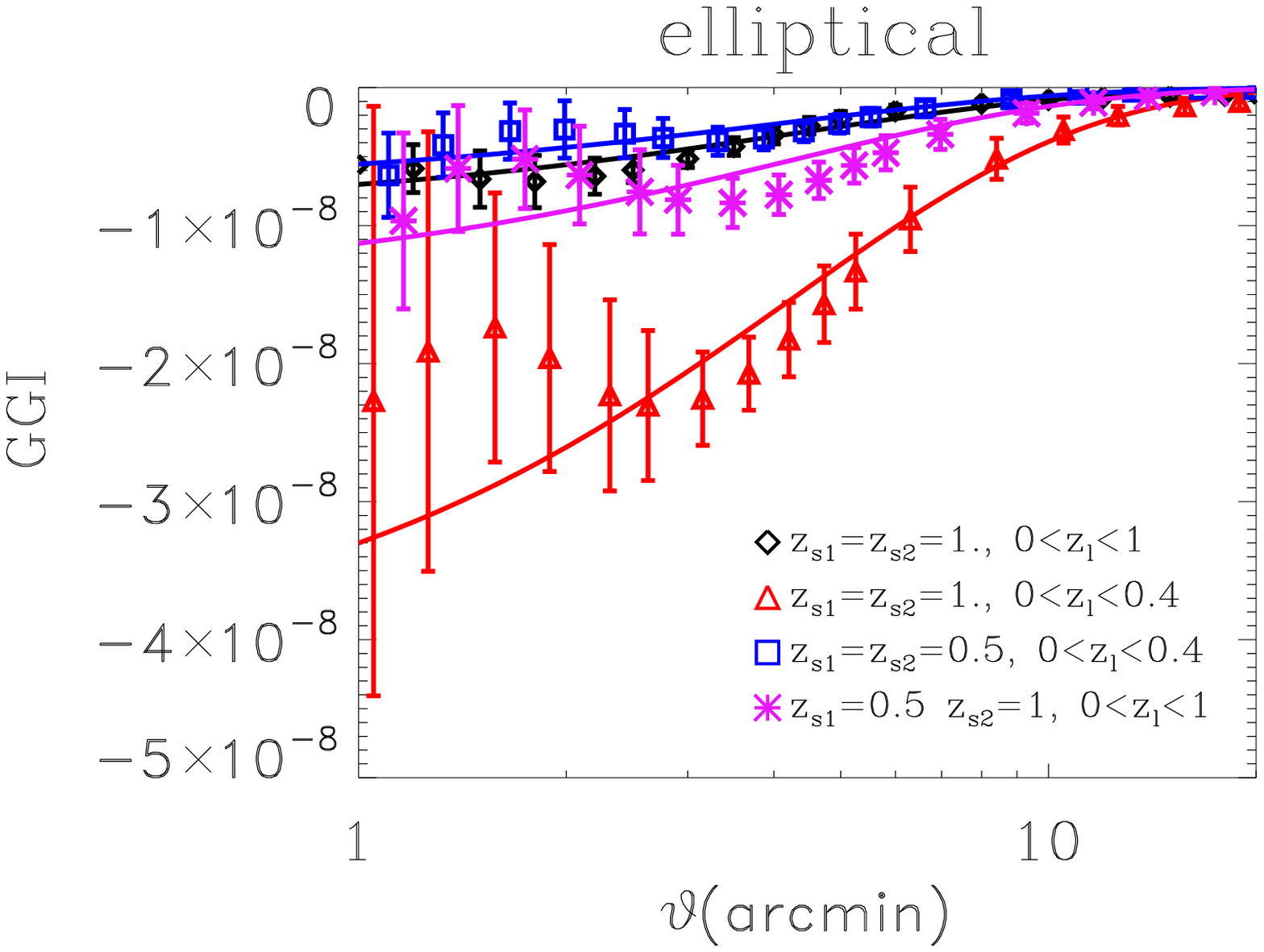,width=.45\textwidth}&\psfig{figure=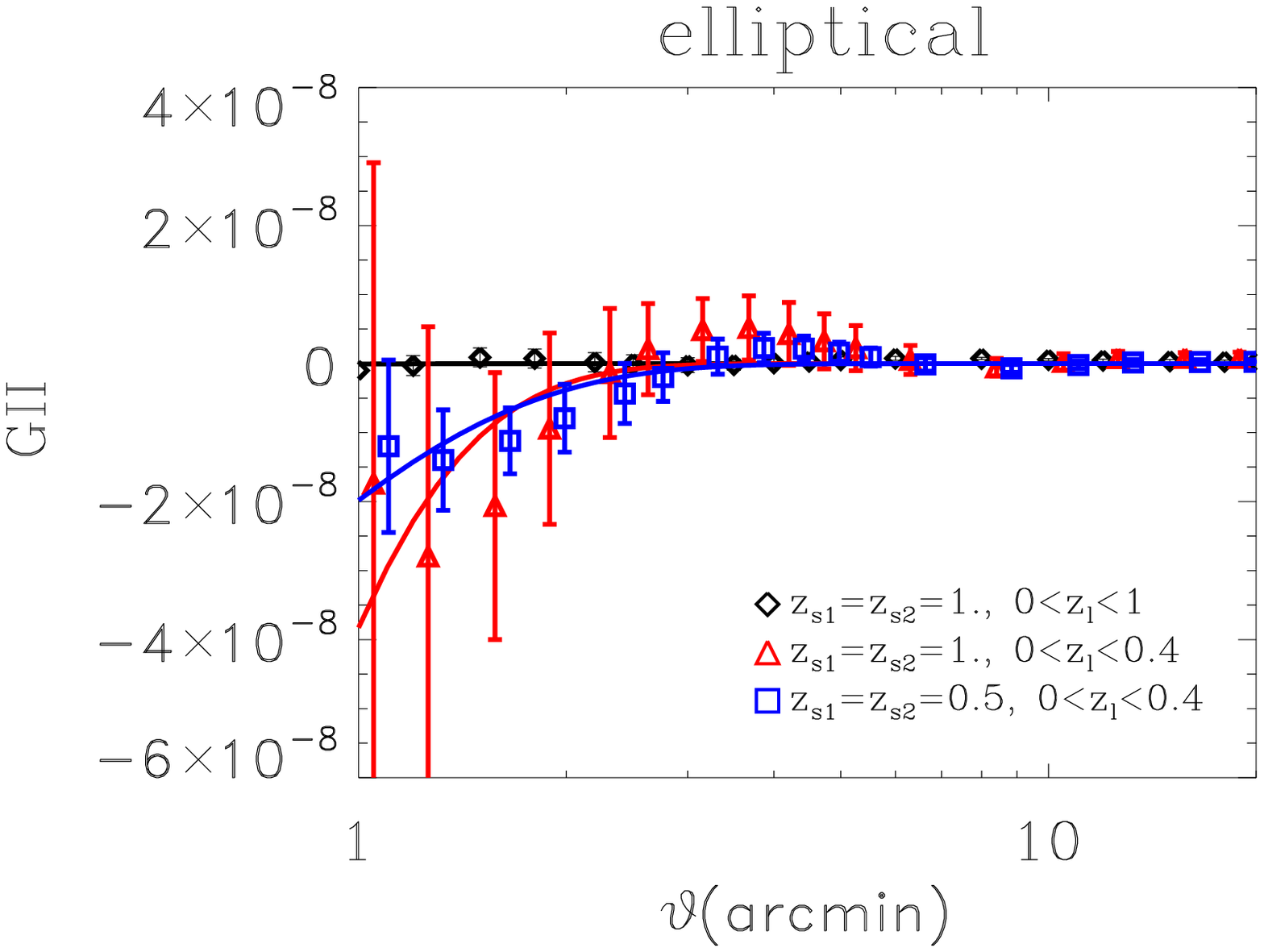,width=.45\textwidth}\\
\psfig{figure=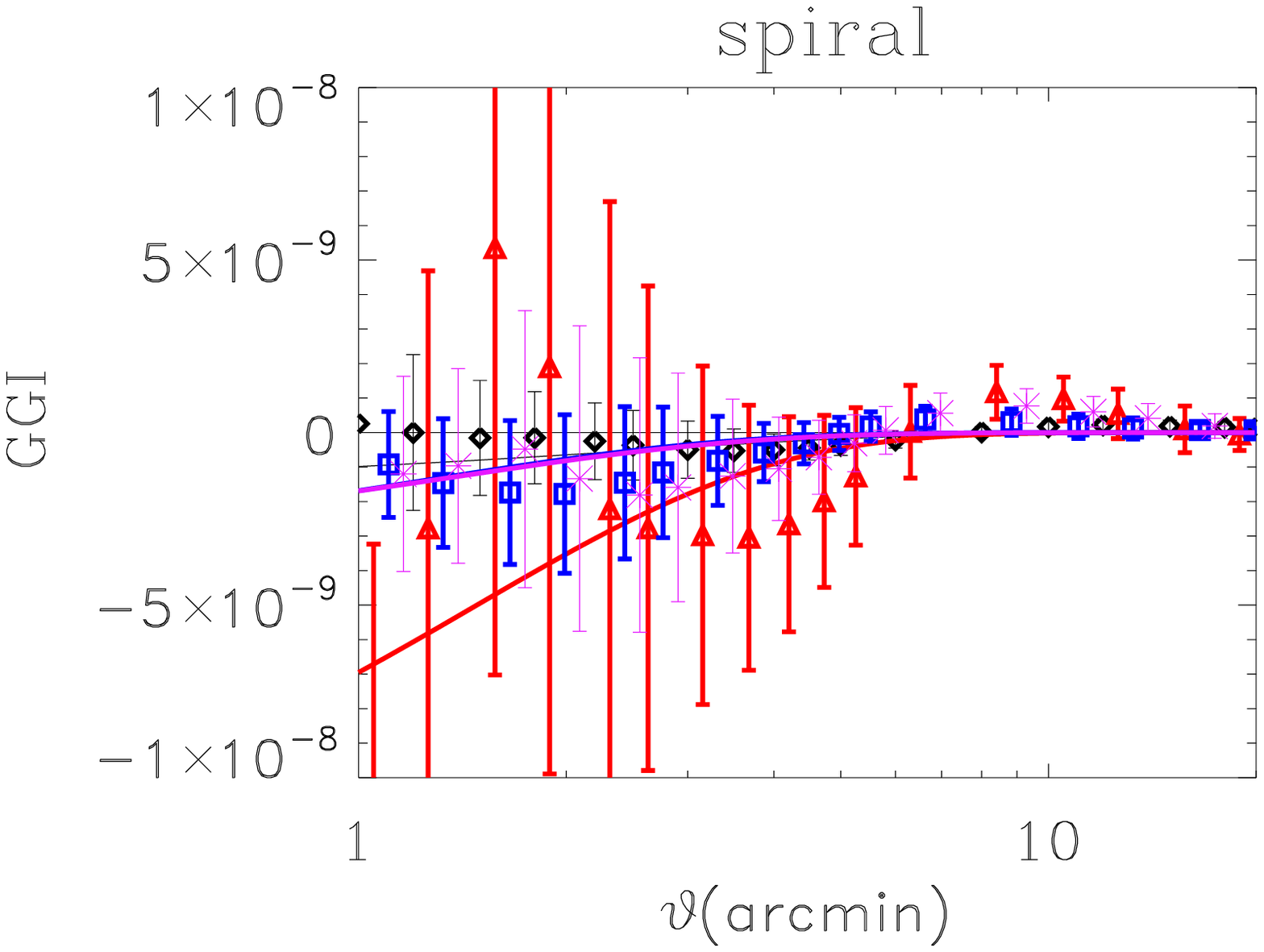,width=.45\textwidth}&\psfig{figure=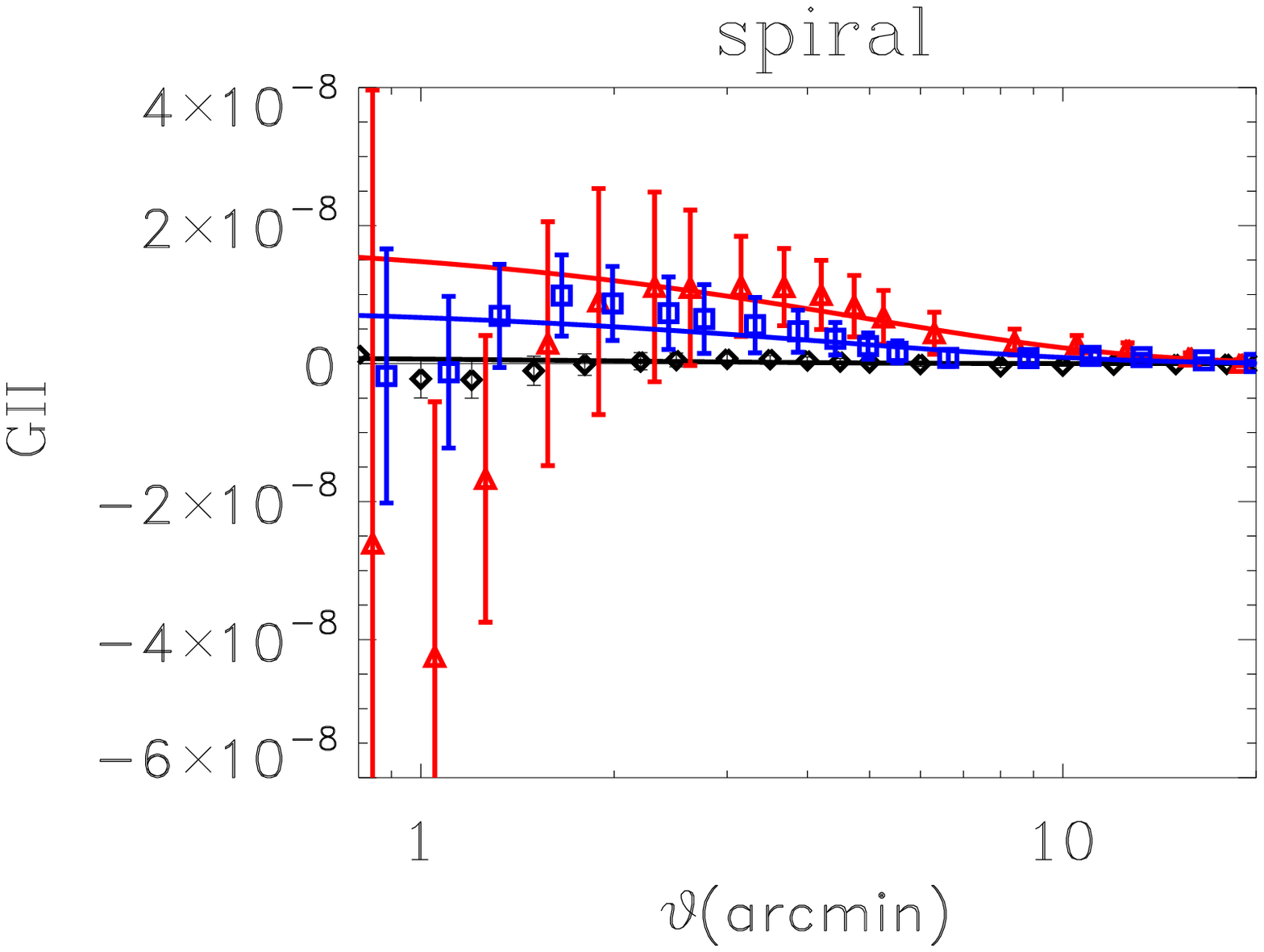,width=.45\textwidth}\\
\psfig{figure=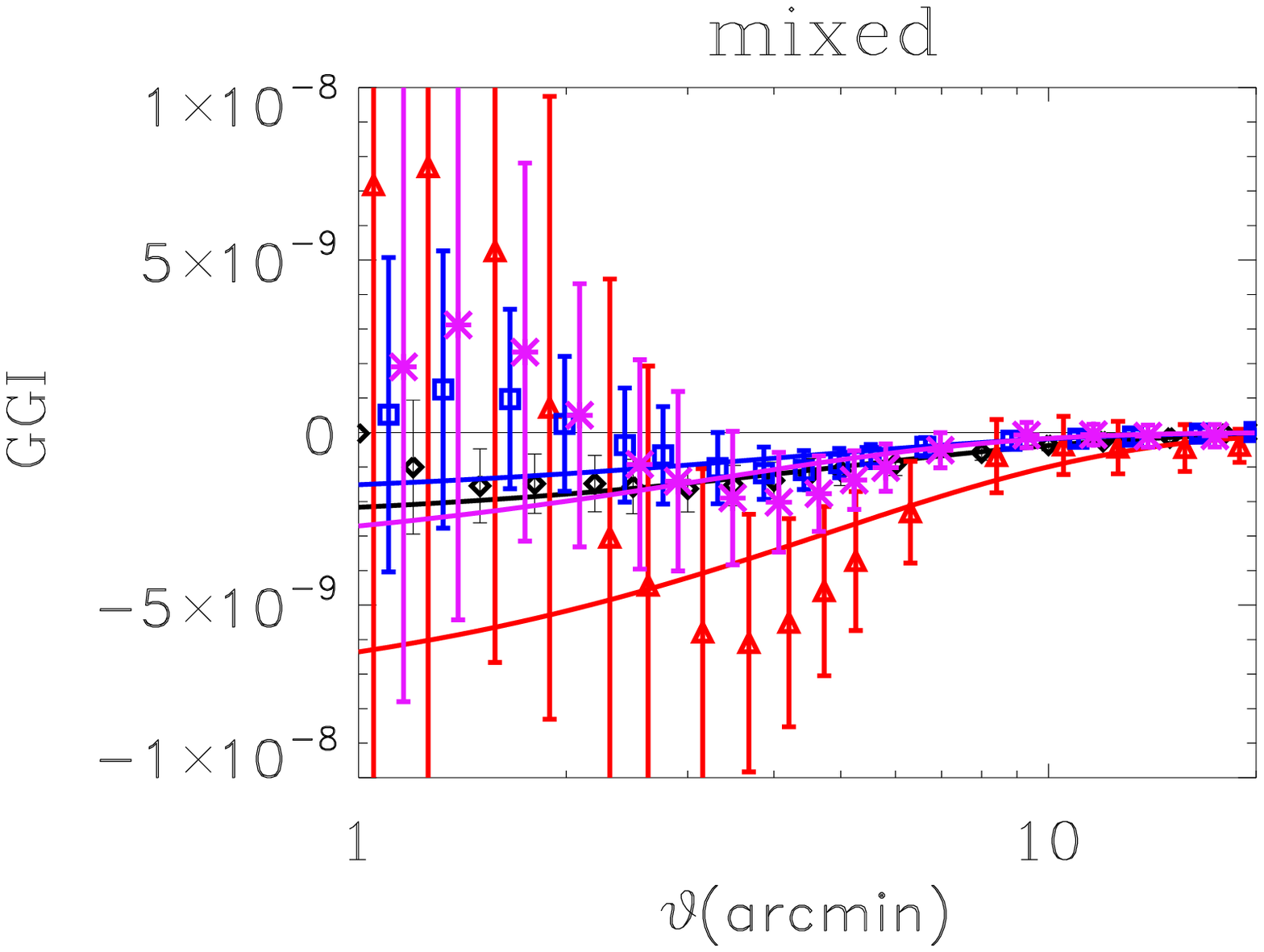,width=.45\textwidth}&\psfig{figure=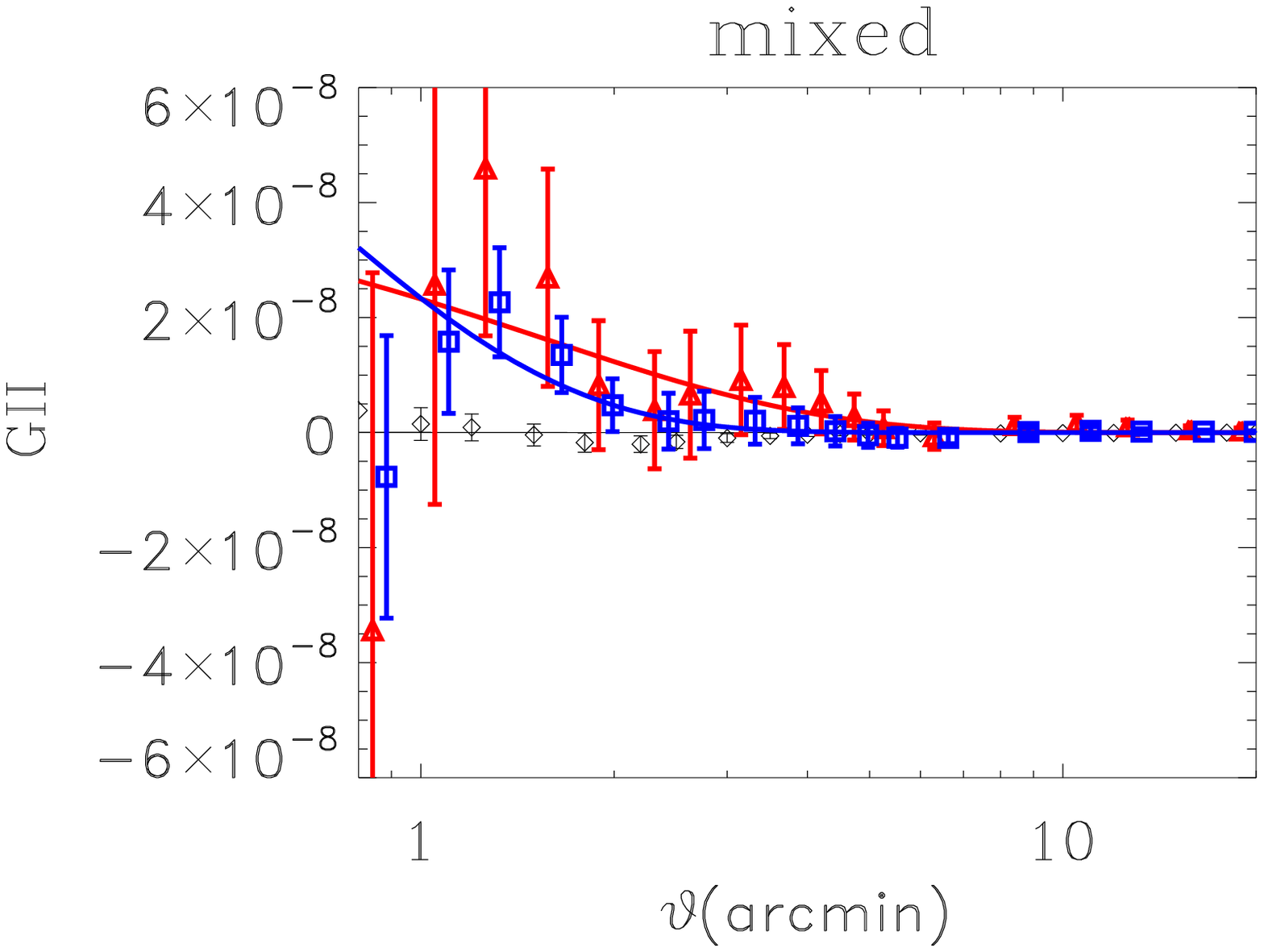,width=.45\textwidth}
\end{tabular}
\caption{\label{SSE} GGI (left panels) and GII (right panels) components for
  different redshift slices and three galaxy models: elliptical (upper
  panels), spiral (middle panels) and mixed (lower panel). We show three sets
  of measurements corresponding to: a lens distribution $0<z_\ll<1$ and a
  source distribution $z_\ss=1.05$ (black diamonds), 
$0<z_\ll<0.4$ and $z_\ss=1.05$
  (red triangles) and $0<z_\ll<0.4$ and $z_\ss=0.45$ (blue squares). 
For the GGI
  component we also include a third model which contains lenses $0<z_\ll<0.4$
  and sources in two different planes:  one at $z_\ss=0.45$ and the other at
  $z_\ss=1.05$. Error bars
  represent the dispersion in the average value between the twelve
  simulations. For each measurement of the GGI and GII components we show  the
  best-fit model (solid lines) obtained by  rescaling the redshift
  distribution  dependence given by 
  equations (\ref{fitmodel})
  and (\ref{fitmodelbis}) with the angular dependence 
described by Equation (\ref{thetamodel}).}
\end{figure*}
In this section we study the behavior of the three-point shear-shape
coupling which, as it has been detailed in 
equations (\ref{GII}) and (\ref{GGI}),
 can be divided in two terms, namely GGI and GII. 
One would expect  these two terms to have a different behavior as a function of the 
redshift distribution of the lenses and of the sources. The GII term depends on the intrinsic ellipticity correlation between two pairs of
galaxies thus it should be  significant only for triplets which  contain
foreground galaxies closer than the scales on which tidal forces act. 
The GGI term  correlates the shape of lensed background
galaxies  with the intrinsic shape of a foreground galaxy so it is expected to be
significant also when correlating well-separated redshift slices, similarly to
what has been found for the GI term (Bridle \& King 2007).
Using both ray tracing simulations with redshift  $z_\ss = 0.45 $ and $z_\ss = 1.05 $ and dividing
the  foreground galaxies into redshift bins, we study the evolution of the GII
and GGI terms as a function of the source and lens redshift
distribution. 
Figure \ref{SSE} shows the amplitude of the GGI (left panels) and GII (right
panels) terms  for elliptical (upper panels), spiral (middle panels) and
the mixed model (lower panels) for  different redshift bins.  
We note that the amplitude  of the GGI and GII terms for a given survey
depends  on the morphology of the galaxies. The elliptical galaxies show  a significant
negative GGI signal whereas the spiral model shows a GGI signal 
which is consistent
with zero. Finally, the mixed model shows a slightly negative GGI signal.  

 The GII component for the spiral galaxies show an
angular dependence similar to the one of the elliptical galaxies and
interestingly  the signal at intermediates scales ($2 \lsim \vartheta \lsim 10$
arcmin) is stronger than for the
 elliptical galaxy model. Similarly to what we previously found for the III term,  the mixed model GII term is different from the
 one obtained using only elliptical or spiral  galaxies.
 This is likely to be a consequence of the fact that  for the two
 morphologies the
 ellipticity depends  differently on the  properties
 of the parent halos, which also could explain the difference in results
 for the  III term. To understand  this phenomena would require building a
 model for the three-point correlation function between weak lensing shear  and tidal
 field, in a similar analysis as that of the two-point correlation function presented in Hirata \&
 Seljak (2004). This however is beyond the scope of this paper. 
 
\begin{figure}
\psfig{figure=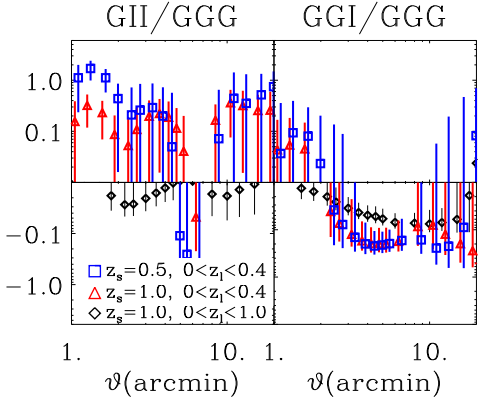,width=0.45\textwidth}
\caption{\label{GGvsGI}  Left side: ratio between the GII term
 and the expected signal (GGG) for three shallow surveys. Right side:
ratio between the GGI term and the GGG  signal for the same surveys as the  left
  panel. The galaxy model is a  mixed model. We used the same surveys as
 Figure \ref{SSE}. The GGG signal has been computed by taking $z_s=1.05$
 (black diamonds and red triangles) and $z_s=0.45$ (blue squares).}
\end{figure}

In order to allow a more quantitative comparison between the weak lensing
three-point  shear signal and the systematics GII and GGI, we  report in
Figure \ref{GGvsGI} the ratio between the terms GII and GGG (left panel) and
GGI and GGG (right panel) for the mixed model.
\begin{table*}
\caption{\label{tab} Summary table of the best fit parameters of the GGI
  (first four columns) and
  GII (last three columns) terms for elliptical (upper lines), spiral (middle
  lines) and mixed (lower lines) galaxies using Equation (\ref{fitmodel}) and Equation (\ref{fitmodelbis}). The values the parameters $\vartheta_0$ 
 are  given in arcmin, the values of the parameter $A$ are given in units of $ 10^{-7} h ~{\rm Mpc}^{-1}$.  
 For each model we report the value of the reduced $\chi^2$. }
\begin{tabular}{|l|l|c|c|c|c|c|c|c|}
&       &          &     GGI  &    &  &    & GII      &                 \\ 
 \hline
&    &$z_\ss=1$&$z_\ss=1$&$z_\ss=0.5$&$z_{\ss}=1;0.5$&$z_\ss=1$&$z_\ss=1$&$z_\ss=0.5$ \\
\hline
&    & $z_\ll<1$  &  $z_\ll<0.4$ & $z_\ll<0.4$   & $z_\ll<0.4$ & $z_\ll<1$  &  $z_\ll<0.4$ & $z_\ll<0.4$   \\
\hline
&$\vartheta_0$&$4.39\pm 0.38$ & $4.23 \pm 0.33 $  & $4.06 \pm 0.45 $   & $ 4.25
 \pm 0.42$  & $0.37\pm 2.02$ & $0.38 \pm 0.34$ & $0.60 \pm 0.23$ \\
elliptical &$A$&  $-0.88 \pm 0.09$ & $-1.02 \pm 0.11$ & $-0.47 \pm 0.07$ & $-0.62
 \pm 0.08$ &   $-0.04 \pm 0.30$ & $-2.56\pm 6.77 $ & $-1.14\pm 1.34$\\
&$\chi^2$ & $1.11$ & $0.59$ & $0.31$ & $0.52$ & $4.34$ & $0.62$ &$0.73$ \\
\hline
&$\vartheta_0$&  $5.12 \pm 0.92$ & $4.84 \pm 1.75$ & $3.24 \pm 1.73$ & $0.38
 \pm 1.73$ &   $0.38 \pm 0.66$ & $1.58\pm 0.70 $ & $0.62\pm 0.28$\\
spiral&$A$&$-0.26\pm 0.05$ & $-0.19 \pm 0.09 $  & $-0.27 \pm 0.07 $   & $ -0.16
 \pm 0.12$  & $0.03\pm 0.11$ & $0.22 \pm 0.19$ & $1.28 \pm 1.37$ \\
&$\chi^2$ & $0.14$ & $0.36$ & $0.23$ & $0.24$ & $0.74$ & $0.48$ &$0.30$ \\
\hline
&$\vartheta_0$&$2.36\pm 1.40$    & $1.46\pm 1.24$ & $1.27\pm 0.83$ & $1.36\pm 1.52$ &$1.94\pm 1.88$&$4.83 \pm 0.90 $&$4.46\pm 0.92$ \\
mixed&$A$&  $-0.15 \pm 0.11$ & $ -0.33 \pm 0.38 $ & $-0.24 \pm 0.21$ & $-0.15 \pm 0.22$ & $0.05\pm 0.07$ & $0.09\pm 0.02$&$0.09\pm 0.03$ \\
&$\chi^2$ &$0.55$ &  $0.51$ &$0.30$ &$0.49$ & $0.29$&$0.55$&$0.40$ \\
\hline

\end{tabular}
\end{table*}
To aid future comparisons we provide fits to our results.  We make the
assumption that for a given triplet  the GGI and GII terms  can be factorized in two functions:
one depending on the comoving distances  of lenses and sources  modeled as in
King (2005) and the other depending on
the angular scale.
We rewrite the GGI term as:
\ba
{\rm GGI}(w_{\rm s_1}, w_{\rm s_2}, w_{\rm L}, \vartheta)={\rm E_{GGI}}(w_{\rm
    s_1}, w_{\rm s_2}, w_{\rm L}) F(\vartheta),
\ea
with:
\ba
{\rm E_{GGI}}(w_{\ss_1}, w_{\ss_2},w_{\rm L})=\int_{0}^{w_{\rm L}} dw_\ll
 \frac{f_k(w_{\ss_1}-w_\ll)}{f_k(w_{\ss_1})} \label{fitmodel}\\
\times\frac{f_k(w_{\ss_2}-w_\ll)}{f_k(w_{\ss_2})} p_\ll(w_\ll)\nonumber, 
\ea
where $w_{\rm L}$ is the maximal comoving distance of the lenses   with the
condition $w_{\rm L} < {\rm min}(w_{\ss_1},w_{\ss_2})$, $f_k(w)$ is the
angular diameter distance, $p_\ll(w_\ll)$ is the
radial lens distribution and $F(\vartheta)$ is a generic function of the angular scale.

Similarly we factorize  GII with the assumption that the intrinsic alignment
acts only for pairs of galaxies  with the same redshift.  Thus we write:
\ba
{\rm GII}(w_{\rm s}, w_{L}, \vartheta)={\rm E_{GII}}(w_{\rm s}, w_{\rm L}) F(\vartheta),
\ea
 with: 
\be
{\rm E_{GII}}(w_{\rm s}, w_{\rm L})=\int_{0}^{w_{\rm L}} dw_\ll \frac{f_k(w_{\ss}-w_\ll)}{f_k(w_{\ss})}  p_\ll(w_\ll)^2\label{fitmodelbis},
\ee
with the condition $w_{\rm L} <w_{\ss}$. 
For a broad source distribution one should integrate Equation (\ref{fitmodel}) and
Equation (\ref{fitmodelbis}) over the source distribution to obtain the average
 effect. However in our case the ray-tracing, source planes 
follow a Dirac distribution
 centered at $z_\ss=0.45$ or $z_\ss=1.05$.
The lens distribution $p_\ll(w_\ll)$ is measured directly from the galaxy
catalogues.
We find that these scaling factors provide a good description of the redshift
 dependence  both for the GII and GGI term.
With this redshift dependence model we try to model the behavior of the  GII and GGI terms as a function
of the angular scale.
We use a two parameter
 function $F(\vartheta)$:
\begin{eqnarray}
F(\vartheta)=A ~ \exp(-\vartheta/\vartheta_0)\label{thetamodel}
\end{eqnarray}
where $A$ and
 $\vartheta_0$ are free parameters.

\begin{figure*}
\begin{tabular}{cc}
\psfig{figure=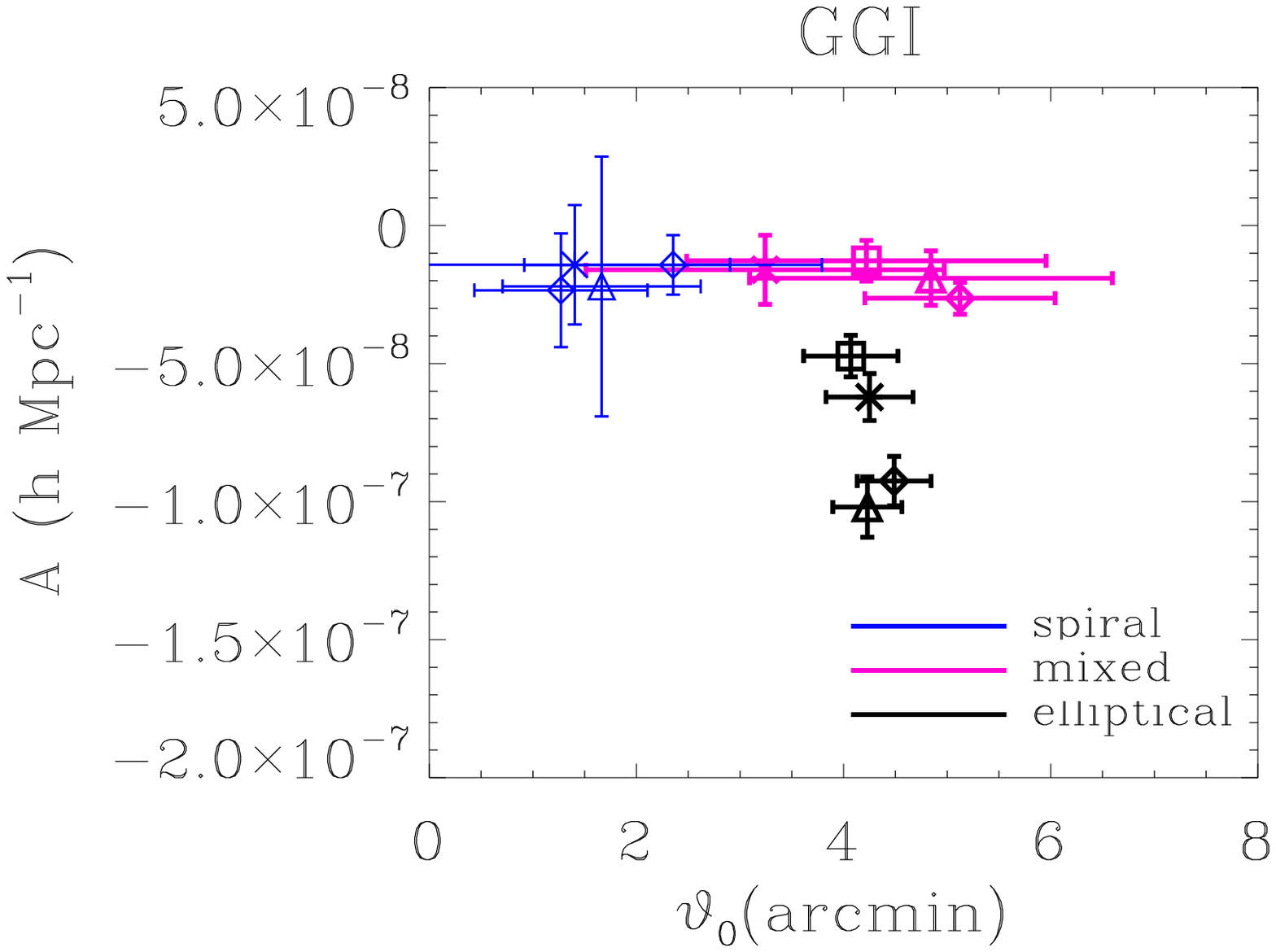,width=.40\textwidth}&\psfig{figure=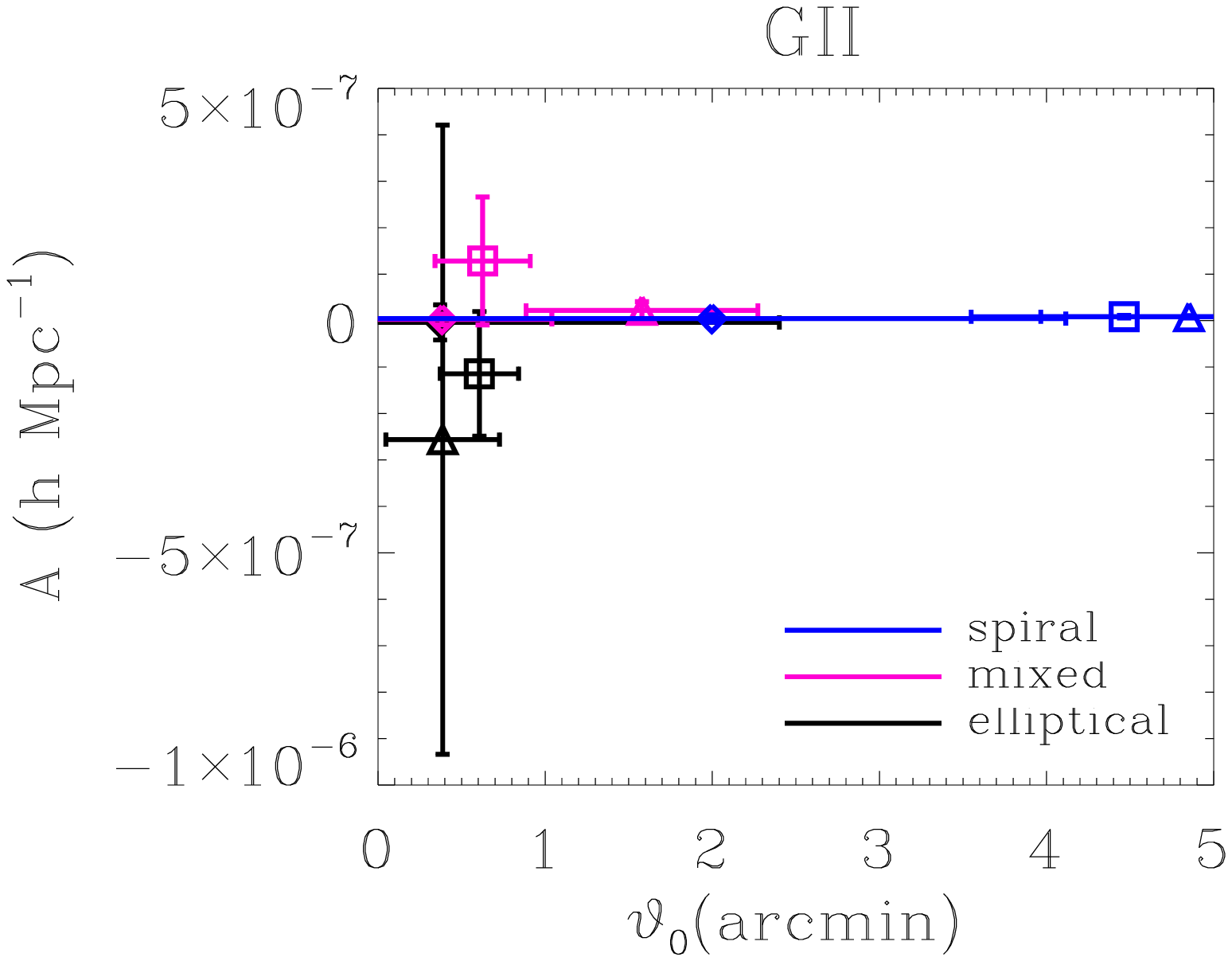,width=.40\textwidth}
\end{tabular}
\caption{ \label{fit}Left panel: best-fit values for  the parameters
  $\vartheta_0$ and $A$ of Equation (\ref{thetamodel}) used to fit the angular dependence of
  GGI for the four redshift distributions; we
  chose to indicate each  source/redshift
  distribution using the  same symbols as Figure \ref{SSE}.  
  Right panel: shows the same parameters
  as left panel but now for the GII component.}
\end{figure*}

In order to avoid bias from the limiting resolution of the simulations we
 chose to perform the fit using only angular scales $\vartheta > 1$ arcmin. 
In Figure \ref{SSE} we compare the measurements of the GGI and GII terms with
 the best-fit model (solid lines) obtained  by using the redshift rescaling
 defined by the equations
 (\ref{fitmodel}) and (\ref{fitmodelbis}) and an angular scale dependence
 defined by Equation (\ref{thetamodel}) for the elliptical, spiral and  mixed galaxies.
Figure \ref{fit} shows the best fit parameters $A$ and $\vartheta_0$ for the GGI (left panel) and the GII (right panel) terms  for several redshift slices, for elliptical, spiral and mixed model. 
These values are also summarized in the table \ref{tab} which includes also the reduced $\chi^2$ of the fit for
both the GGI and GII terms. The small values of the $\chi^2$ show that
 the model we suggest is a good fit to the measured GGI
 component. However, for the elliptical model the dispersion of the best-fit
 $A$ parameter between  the four redshift slices (see left panel of Figure
 \ref{fit}) is larger than the error bars. This suggests that the
 redshift-dependent rescaling described by the Equation (\ref{fitmodel}) could 
require some modification. 
The GII component  is generally more noisy than the GGI component,  
and for  this reason it is
 hard to establish whether the model defined by equations (\ref{fitmodelbis}) and
 (\ref{thetamodel}) is a good one.
However, for the mixed model, which is our most realistic model, the GII term
 can be fairly well described by our fitting function.
Increasing the size of
simulations in future analyses will allow us to improve upon these models.

\begin{figure*}
\begin{tabular}{cc}
\psfig{figure=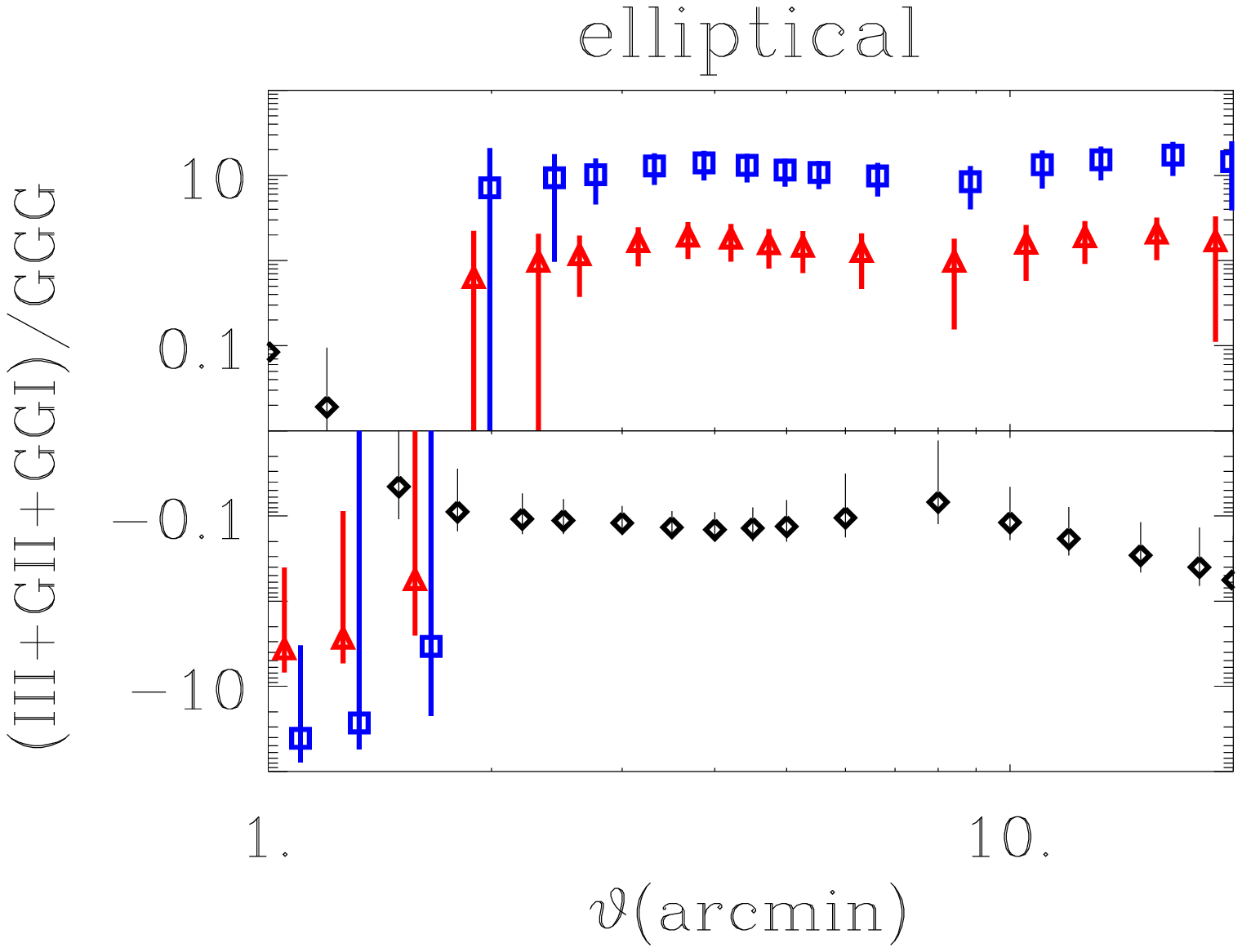,width=.45\textwidth}&\psfig{figure=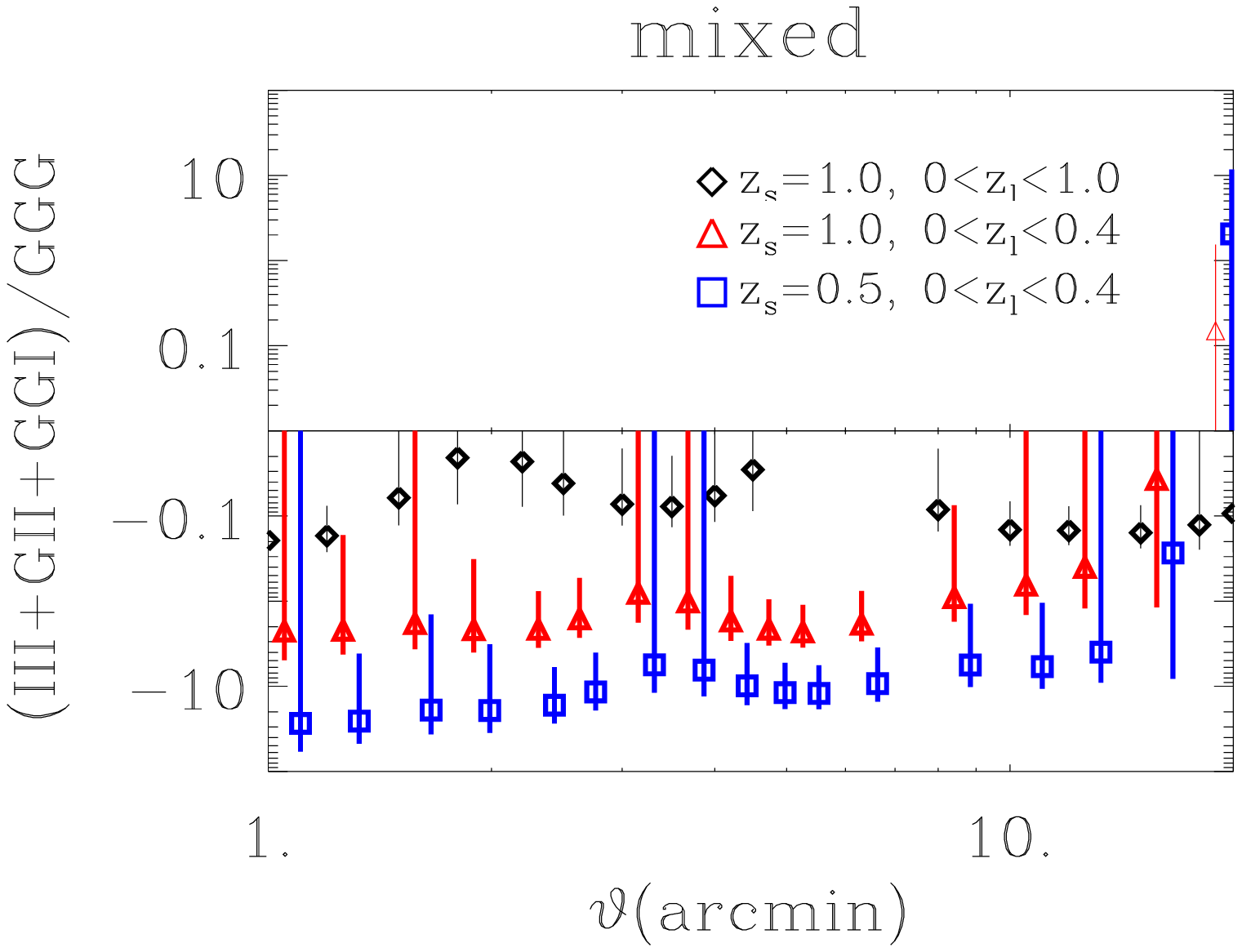,width=.45\textwidth}
\end{tabular}
\caption{\label{SEE_sys} Ratio between   the sum of the terms GII, GGI and III
  and the weak lensing third order moment signal GGG, for the elliptical (left panel) and mixed model
  (right panel). Three cases are shown: $z_\ss=1$, $0<z_\ll<1$
  (black diamonds), $z_\ss=1$, $0<z_\ll<0.4$ (red triangles) and  $z_\ss=0.5$, $0<z_\ll<0.4$ (blue squares).  For the first and second case the GGG component is
  computed assuming $z_\ss=1.05$. For the third case $z_\ss=0.45$.}
\end{figure*}

Because of  the different dependence on the morphology of galaxies and on the redshift
distribution of the sources and lenses for the III, GGI and GII terms, one
may be interested in knowing the total effect of the intrinsic
alignments and of the shear-coupling on the three-point shear for realistic surveys.
 Figure \ref{SEE_sys}, shows the ratio between the sum of  GGI, GII and III terms
 and  the expected weak lensing  signal for several redshift distributions
 for both the elliptical (left panel) and mixed (right panel) model.

 For shallow surveys ($z_\ll \lsim 0.5$), the ratio is largely dominated by the III term.
This is true both for elliptical and mixed model with the difference that for
the elliptical  galaxies the intrinsic alignment enhances the weak lensing
shear signal whereas
for the mixed model it suppresses the weak lensing signal.
For a deep survey $z_\mm \simeq 0.7$  the ratio is dominated by the
GGI term, whose amplitude is $10\%$ of the GGG signal for elliptical galaxies
and few percent for  mixed model.

\begin{figure}
\psfig{figure=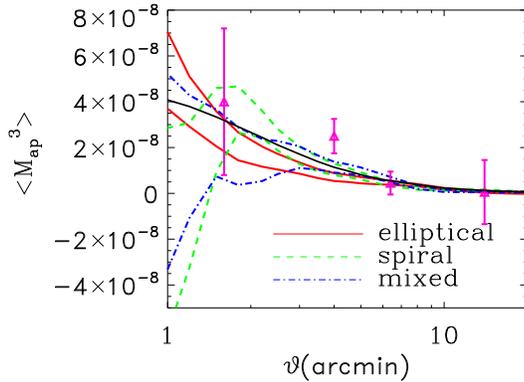,width=.45\textwidth}
\caption{\label{Jarvis}Third order moment of the aperture mass  $\langle M_{\rm
    ap}^3\rangle_\vartheta$ measured by Jarvis et al. (2004) on the CTIO data (pink
    triangles) compared with the expected measured  third order aperture mass
    statistics, i.e. GGG+III+GII+GGI, for the three models of galaxies used in
    this paper. For each model we show a lower  and upper value of the
    total third order moment  given  by  the average  $\pm 1
    \sigma$ error bars. This error is
    computed as a quadratic sum of the error affecting each term.
The GGG weak lensing signal is the same used in Jarvis et al. (2004), i.e. the
    one for a single source redshift at $z_\ss=0.66$. The III signal has been computed using a galaxy redshift
    distribution with $z_\ll<0.8$ which is characterized by the  similar median
    redshift $z_\mm \sim 0.66$ used to compute the GGG signal. The GII and GGI terms have been rescaled by
    using the equations (\ref{fitmodel}) and (\ref{fitmodelbis}) for a survey
    with $z_\ss=0.66$ and $z_\ll < 0.66$. For comparison we plot the expected weak
    lensing  $\langle M_{\rm ap}^3\rangle_\vartheta$ (black solid line). }
\end{figure}

In order to compare our results to observations we present in
in Figure \ref{Jarvis} the  $\langle M_{\rm ap} ^3 \rangle $ 
results by Jarvis et al. (2004) and the 
$ \langle M_{\rm ap}^3 \rangle $ that we would expect for this 
same survey from weak gravitational lensing and the III, GGI and GII terms. 
For this comparison we have used all of the three galaxy models.
We compute the GGG signal using a redshift plane at $z_\ss=0.66$  
as done by 
Jarvis et al. (2004). We  add the III model using  galaxies
with  $z<0.8$ so that the  redshift distribution is characterized by the
a  median redshift  similar to the one of the CTIO galaxy catalogue. We use Equation (\ref{fitmodel}) 
and Equation (\ref{fitmodelbis})
to rescale the GII and GGI to a survey with $z_\ss=0.66$ and $z_\ll<0.66$.  
For this survey  we find that the dominant contribution is given by the 
GGI which slightly decreases the estimated the
weak lensing signal.  We find that all 
three galaxy models are consistent with the Jarvis et al. (2004) results.

\section{Conclusion}
Using a set of realistic N-body $\Lambda$CDM simulations we have explored the effect of intrinsic galaxy alignments and the coupling between weak lensing and the foreground  ellipticity field  on
the third order moment  of the aperture mass. 
We find that the intrinsic alignment
dominates the three-point shear signal for shallow
surveys.  For deeper surveys
the intrinsic alignment is less significant, as a result of projection
effects. Nevertheless, if not taken into account properly, it will still limit the
accuracy of tomography measurement and affect  the
predictions of cosmological values for the next generation of 
very large surveys, where the signal to
noise ratio is high.

 We found that  for a given survey
depth intrinsic alignments affect the three-point weak lensing statistics more strongly than the two-point
shear statistics. In other words, in order to achieve the same level of
accuracy in the three-point shear statistics measurement as the two-point weak
lensing shear statistics one needs deeper surveys. 

Overall, this result shows once more the importance of the knowledge of redshift for
each individual galaxy for high-precision cosmology.
Knowing the
redshift  of each source allows one to remove the bias from  intrinsic galaxy 
alignments  by discarding physically close
triplets (pairs) when computing the three-point (two-point) shear statistics.
Moreover the knowledge of redshift allows one to model the intrinsic alignment
and remove its effect on the two-point weak lensing shear statistics (Joachimi
\& Schneider, in prep.).

In this perspective, the measurement of the three-point shear statistics from a
shallow survey offers a particularly effective way to test intrinsic alignment
models. As we showed in this work, the three-point shear statistics in
low redshift  bins are dominated by the intrinsic alignment term, permitting accurate measurements on intrinsic alignment models. As  future dark energy surveys will rely heavily on the good modeling of these effects so they can be marginalised out, this result provides an important new route to constrain and model this physical systematic effect. 
It is indeed possible to choose low-redshift
bins in order to enhance the
intrinsic alignment signal so that the shear signal becomes negligible;
 for example, by selecting galaxies at $z<0.4$.  In this case  
the III/GGG ratio is around 
fifty, allowing one to study the intrinsic alignment, essentially
without contamination from the weak lensing shear.

Using projected mass maps we have 
studied the shear-shape coupling effect which  is
also likely to bias the measurement of the third order moment of weak lensing shear. 
We showed that this systematic can be described
by two terms, GGI and GII, which affects the third order moment measurement 
in a non-trivial  way and is
dependent  on the distance  between  the lenses and sources  and on the
morphology of galaxies. 
Even for a moderately deep survey like the
CFHTLS Wide the amplitude of the third order moment of the shear estimation could be
underestimated by $\sim 5-10\%$. 
For shallower surveys such as KIDS or PanSTARRS-1 
the bias  is expected to be higher.  Our results show that it will not be possible to carry out precise three-point cosmic shear measurements with these surveys without modeling the coupling between weak lensing shear and
intrinsic alignment.
Also for the next generation of deep large surveys, such as SNAP, DUNE or LSST 
the precision one can
achieve on the cosmological constraints relies in the ability to model 
and marginalize out the
shear-shape and intrinsic correlations. 
     
It is possible to use  simple models to
describe both the angular and redshift dependence of the GGI and
GII terms. These models can be used to correct the effect of the
coupling between shear and intrinsic alignment on the third order moment of the aperture
mass.

Unfortunately, the sample available for this work is not big enough to give
a definitive answer; more precisely, the parametric models we used to fit the GGI and GII
components are marginally constrained, due to the large statistical and sampling
variance affecting our sample. A more detailed study  of the redshift
dependence of the shear-ellipticity correlation will be  required. Such a study should use a larger sample of simulations in order to better
 constrain our models. Furthermore, we still need to develop a method to include   satellite galaxies and  account for environmental effects when
 determining the ellipticity of galaxies in a single parent halo.  In addition, 
 an improved mixed galaxy population as a function of redshift evolution would
 also play an important role in establishing the correct average effects, as we
 have shown in this paper the net effect depends on the morphology of the
 galaxy population.
 We also think that a comparison between results from simulations with
  results from  
  real data is the best way
  to validate our results. Concerning this point, the agreement between simulations and real data 
shown in the study of the II (H06) and GI (Hirata et al. 2007) components is a
  good indication of the fact that the simulation used in this work, even if it
  represents a simplified model, shows the correct dependence of the intrinsic
  alignment effect on morphology of the galaxies, luminosity  and redshift.

Like other previous works on intrinsic alignment and shear-shape
coupling, this paper demonstrates  the great importance of reliable redshifts 
for future and
current galaxies surveys, which could be used, in parallel with simulated
catalogues,  to study and correct for
intrinsic alignment and shear-shape coupling on the two- and three-point weak
lensing statistics.

\section{Acknowledgments}
We thank  Alan Heavens and Ismael Tereno for helpful suggestions on this
project and Asantha
Cooray for making code based on the CLF  publicly  available.
We also would like to thank Martin White for  generating the N-body
simulations used for this work. 
These simulations were performed on the IBM-SP at NERSC.
ES is supported by the Humboldt Foundation.  
CH is supported by the European Commission Programme in the framework of a Marie Curie Fellowship under contract MOIF-CT-2006-21891.
LW acknowledges  support from  NSERC and CIAR.  
This work has been supported by the RTN Network DUEL and the DFG through the
TR33 `The Dark Universe' and the project SCHN 342/6--1.

\end{document}